\documentclass[aps,showpacs,preprintnumbers,amsmath,amssymb]{revtex4}
\usepackage{dcolumn}
\usepackage[dvips]{epsfig}
\usepackage{graphicx}
\usepackage{amssymb}
\usepackage{color}
\usepackage{enumerate}
\usepackage{subfigure}
\usepackage{appendix}

\begin{document}
\baselineskip=0.8 cm
\title{\bf Electromagnetic emissions from near-horizon region of an extreme Kerr-Taub-NUT Black hole}

\author{Fen Long$^{1}$,  Songbai Chen$^{1, 2}$\footnote{Corresponding author: csb3752@hunnu.edu.cn}, Jieci Wang$^{1}$ \footnote{jcwang@hunnu.edu.cn},  Jiliang
Jing$^{1, 2}$ \footnote{jljing@hunnu.edu.cn}}

\affiliation{$ ^1$ Department of Physics, Key Laboratory of Low Dimensional Quantum Structures
and Quantum Control of Ministry of Education, and Synergetic Innovation Center for Quantum Effects and Applications,
Hunan Normal University, Changsha, Hunan 410081, People's Republic
of China\\
$ ^2$Center for Gravitation and Cosmology, College of Physical Science and Technology,
Yangzhou University, Yangzhou 225009, People's Republic
of China}

\begin{abstract}
\baselineskip=0.6 cm
\begin{center}
{\bf Abstract}
\end{center}

 We have studied electromagnetic line emissions from near-horizon region in the extremal Kerr-Taub-NUT black hole spacetime and then probe the effects of NUT charge on the electromagnetic line emissions.  Due to the presence of the NUT charge,  the equatorial plane is no more a symmetry plane of the KTN spacetime, which leads to that the dependence of electromagnetic line emission on the NUT charge for the observer in the Southern Hemisphere differs from that in the Northern one. Our result indicate that the electromagnetic line emission in the Kerr-Taub-NUT black hole case is brighter than that in the case of Kerr black hole for the observer in the equatorial plane or in the Southern Hemisphere, but it becomes more faint as the observer's position deviates far from the equatorial plane in the Northern one. Moreover, we also probe effects of redshift factor on electromagnetic emission from  near-horizon region in the extremal Kerr-Taub-NUT black hole spacetime.

\end{abstract}
\pacs{ 04.70.Dy, 95.30.Sf, 97.60.Lf }

 \maketitle
\newpage
\section{Introduction}

The existence of black hole in our Universe is confirmed exactly by the gravitational waves detected by LIGO \cite{P1,P2,P3,P4,P5,P6}, which triggers a new era in
observing black hole astrophysics. Another important experiment of observing astrophysical black hole is
the Event Horizon Telescope, which could capture the first image of the supermassive black hole at the center of our Galaxy \cite{P7,P8,P9,P10,P11,P12}, and then provide us a great deal of  information in the near-horizon region of the black hole. With these characteristic information, one can identify the black hole parameters and examine theories of gravity. For a rapidly rotating Kerr black hole, one of such kind characteristic information is the so-called ``Near-Horizon Extremal Kerr line (NHEK line)" , which is a vertical line segment on the edge of the shadow of the high spinning black hole \cite{P15,P16,P17,P18}. A. Lupsasca \textit{etal} \cite{P19} investigated the universal feature of the observed flux at spatial infinity  for the electromagnetic line emissions from the NHEK line, which is emitted by the particles in the innermost part of a radiant thin accretion disk around a high-spin black hole. Their result indicate that the luminosity of electromagnetic line emission for observer is the brightest as both the sources and observer lie in the equatorial plane.
Moreover, they also discuss the change of the flux with the redshift factor. These properties of electromagnetic line emissions could be applied to analyze the behavior of FeK$\alpha$ line emissions from high-spin black holes observed by the experiments, such as XMM-Newton, Suzaku, and NuSTAR \cite{obs1,obs2,obs3,obs4}. Obviously, the properties of the electromagnetic line emissions depend on black hole parameters. In order to understand features of black hole and to examine further the no-hair theorem, it is very significant to study in detail the properties of these near-horizon electromagnetic emissions from other rapidly rotating black holes in different theories of gravity.

In general relativity, another important stationary and axisymmetric metric is Kerr-Taub-NUT (KTN) metric, which is a solution of Einstein field equations  with gravitomagnetic monopole and dipole moments \cite{KTN1,KTN2}. Besides the mass $M$ and the spinning parameter $a$, the KTN spacetime own an extra NUT charge, $n$, which  plays the role of a magnetic mass inducing a topology in the Euclidean section.
Comparing with Kerr spacetime, KTN spacetime with gravitomagnetic monopole
is asymptotically non-flat due to existence of the NUT charge.  Although the
KTN has no curvature singularities, there exist conical singularities on the axis of symmetry. One can
remove conical singularities by imposing a periodicity condition on  the time coordinate \cite{tnnc}.
However, the periodicity condition inevitably yields the appearance of closed timelike curves (CTCs) in the spacetime and then the causality is violated. Moreover,
it also leads to that the analytically extended Taub-NUT spacetime is geodesically incomplete \cite{tnnc1}. On the other hand, Bonnor {\textit et al} \cite{KTNapp1,tnn0,tnn1,tnn2} gave up the time periodicity condition and interpreted the conical singularities as ``a linear source of pure angular momentum", which preserves the causality in the spacetime.   Without the time periodicity condition, Miller {\textit et al} \cite{tnn3} indicated that the vacuum Taub-NUT spacetime can be extended maximally
through both horizons and the related analysis of the geodesic has also been investigated in this spacetime \cite{tnn3,tnn4}.
Despite some unpleasing physical features, the KTN spacetime  is attractive for exploring various physical phenomena in general relativity \cite{KTNapp2,KTNapp3,KTNapp4,KTNapp5,KTNapp6,KTNapp7,KTNapp8,KTNapp9,P23,tdk1}. For example,
the size of the black hole shadow is found to be increase with the NUT charge \cite{KTNapp8}. Moreover,  the particle acceleration  has also been investigated in the background of the KTN spacetime \cite{KTNapp7}, which shows that the NUT charge modifies the restrict
conditions of the rotation parameter $a$ when the arbitrarily high center-of-mass energy appears in
the collision of two particles. The effects of the NUT charge on the motion of photon, the inner-most stable circular orbits and the perfect fluid disk around the black hole have also been studied in Refs.\cite{P23,tdk1}, respectively.  These effects originating from NUT parameter
may help us to probe the gravitomagnetic masses in astronomical observation in the future.

For a high spinning KTN black hole, it is well known that there also exists a vertical line segment on the edge of the black hole shadow \cite{KTNapp8}, which can be called
``Near-Horizon Extremal KTN line (NHEKTN line)" as in the case of Kerr black hole. Therefore, it is natural to ask what features the electromagnetic emissions from the NHEKTN line possess in the spacetime with NUT charge. So the purpose of this paper is to study the effects of the NUT charge on the electromagnetic line emissions from the NHEKTNline.

This paper is organized as follows. In Sec.II, we adopt to the method used in Ref.\cite{P19} and present the formula of the flux for the electromagnetic line emissions from near-horizon region in the extremal KTN black hole spacetime. In Sec.III, we analyze numerically the effects of NUT charge on the electromagnetic line emissions from near-horizon region in the extremal KTN background.
Finally, we present results and a brief summary.

\section{Electromagnetic line emissions from near-horizon region of an extreme Kerr-Taub-NUT black hole}

KTN spacetime is a stationary, axisymmetric vacuum solution of Einstein equation, which describes the gravity of a rotating source equipped with a gravitomagnetic monopole moment. The line element for the KTN spacetime, in the Boyer-Lindquist coordinates, can be expressed as \cite{KTN1,KTN2}
\begin{eqnarray}
ds^2&=&g_{tt}dt^2+g_{\hat{r}\hat{r}}d\hat{r}^2+g_{\theta\theta}d\theta^2+g_{\phi\phi}d\phi^2
+2g_{t\phi}dtd\phi,
 \label{metric1}
\end{eqnarray}
with
\begin{eqnarray}
&&g_{tt}=-\frac{\Delta-a^2\sin^2\theta}{\Sigma},\;\;\;\;\;\;\;\;\;\;\;\;\;\;\;\;\;\;\;\;\;\;
g_{t\phi}=\frac{\Delta \chi-a(\Sigma+a\chi)\sin^2\theta}{\Sigma},\nonumber
\\
&&g_{\phi\phi}=\frac{(\Sigma+a\chi)^2\sin^2\theta-\chi^2 \Delta}{\Sigma},\nonumber\\
&&g_{\hat{r}\hat{r}}=\frac{\Sigma}{\Delta},\;\;\;\;\;\;\;\;\;\;\;\;\;\;\;\;
g_{\theta\theta}=\Sigma,\;\;\;\;\;\;\;\;\;\;\;\;\;\;\;\;\Sigma=\hat{r}^2+(\hat{n}+a\cos\theta)^2,\nonumber\\
&&\Delta=\hat{r}^2-2M\hat{r}+a^2-\hat{n}^2,
 \;\;\;\;\;\;\;\;\;\;\;\;\;\;\;\;\chi=a\sin^2\theta-2\hat{n}\cos\theta,
\end{eqnarray}
where $M$, $a$, and $\hat{n}$ are the mass, the rotation parameter, and the NUT charge of the source, respectively. The NUT charge $\hat{n}$ describes the strength of gravitomagnetic monopole, which is the gravitational analogue of a magnetic monopole in Maxwell's electrodynamics.
As $\hat{n}$ vanishes, the metric (\ref{metric1}) reduces to that of the usual Kerr spacetime. The positions of event horizon and Cauchy horizon of the KTN spacetime are located at
\begin{eqnarray}
\hat{r}_{H,C}=M\pm\sqrt{M^2-a^2+\hat{n}^2},
\end{eqnarray}
 which are the roots of the equation
\begin{eqnarray}
\hat{r}^2-2M \hat{r}+a^2-\hat{n}^2=0.
\end{eqnarray}
With increase of the NUT parameter $\hat{n}$, it is obvious that the radius of the event
horizon becomes larger. As $M^2-a^2+\hat{n}^2=0$, these two horizons emerge and then the KTN black hole turns to be extremal in this case.

In order to study electromagnetic emissions from near-horizon region of an extremal Kerr black hole,  Lupsasca {\textit {et al}} \cite{P19} suppose a thin stationary and axisymmetric accretion disk around the black hole in which the timelike particles move along the circular orbit in the equatorial plane, and then assume that all of electromagnetic radiation caused by the source propagate along null geodesic from the disk to the distant observer located in the position with the polar coordinate $r_0$ and the polar angle $\theta_0$. However, as in the previous discussion, the presence of the NUT charge leads to that the closed timelike curves could appear in the KTN spacetime, which implies that the method adopted in Kerr spacetime could not be valid in the regions with closed timelike curves in KTN spacetime. Thus, it is necessary to pick out the regions with no any closed timelike curves in which one can use the same method as in Kerr case.
Here, we adopt Bonnor's viewpoints {\textit et al} \cite{KTNapp1,tnn0,tnn1,tnn2} and gave up the time periodicity condition, which avoids the closed timelike curves originating from the periodicity of time coordinate. However, there could exist another type of closed timelike curves in KTN spacetime, which is caused by the periodicity of angular coordinate under certain conditions.
\begin{figure}
\center
\includegraphics[width=5cm]{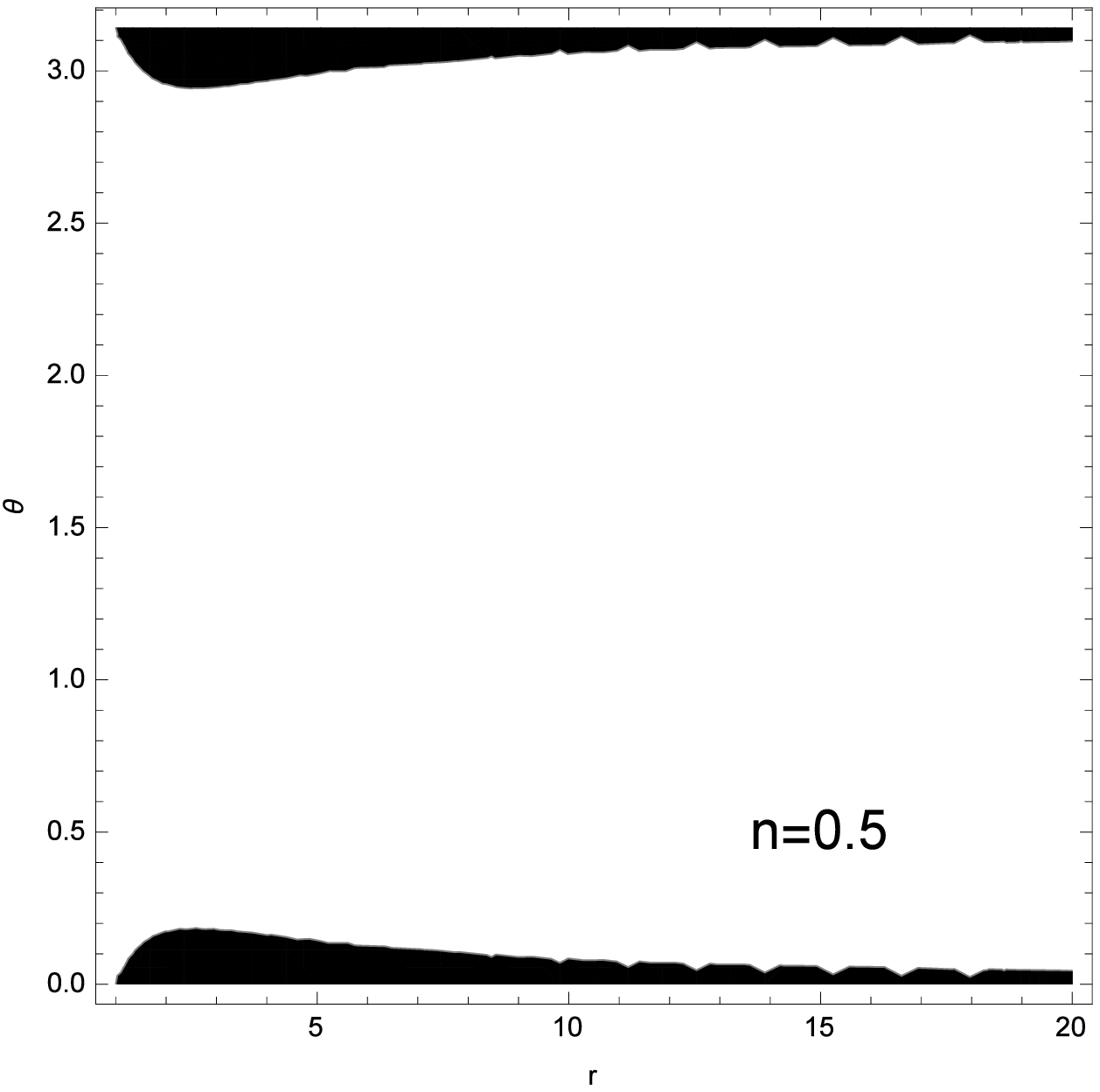}\includegraphics[width=5cm]{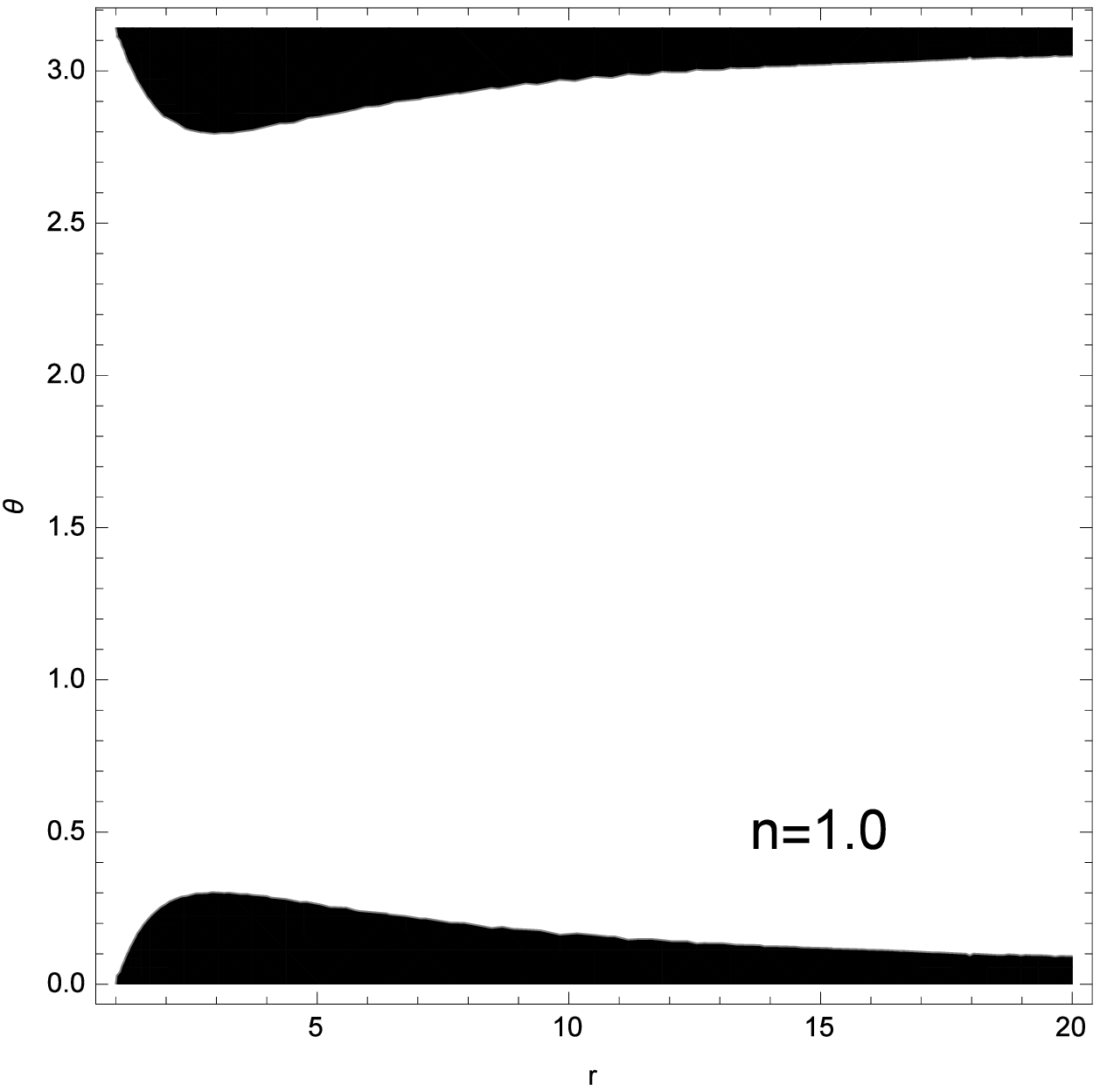}
\includegraphics[width=5cm]{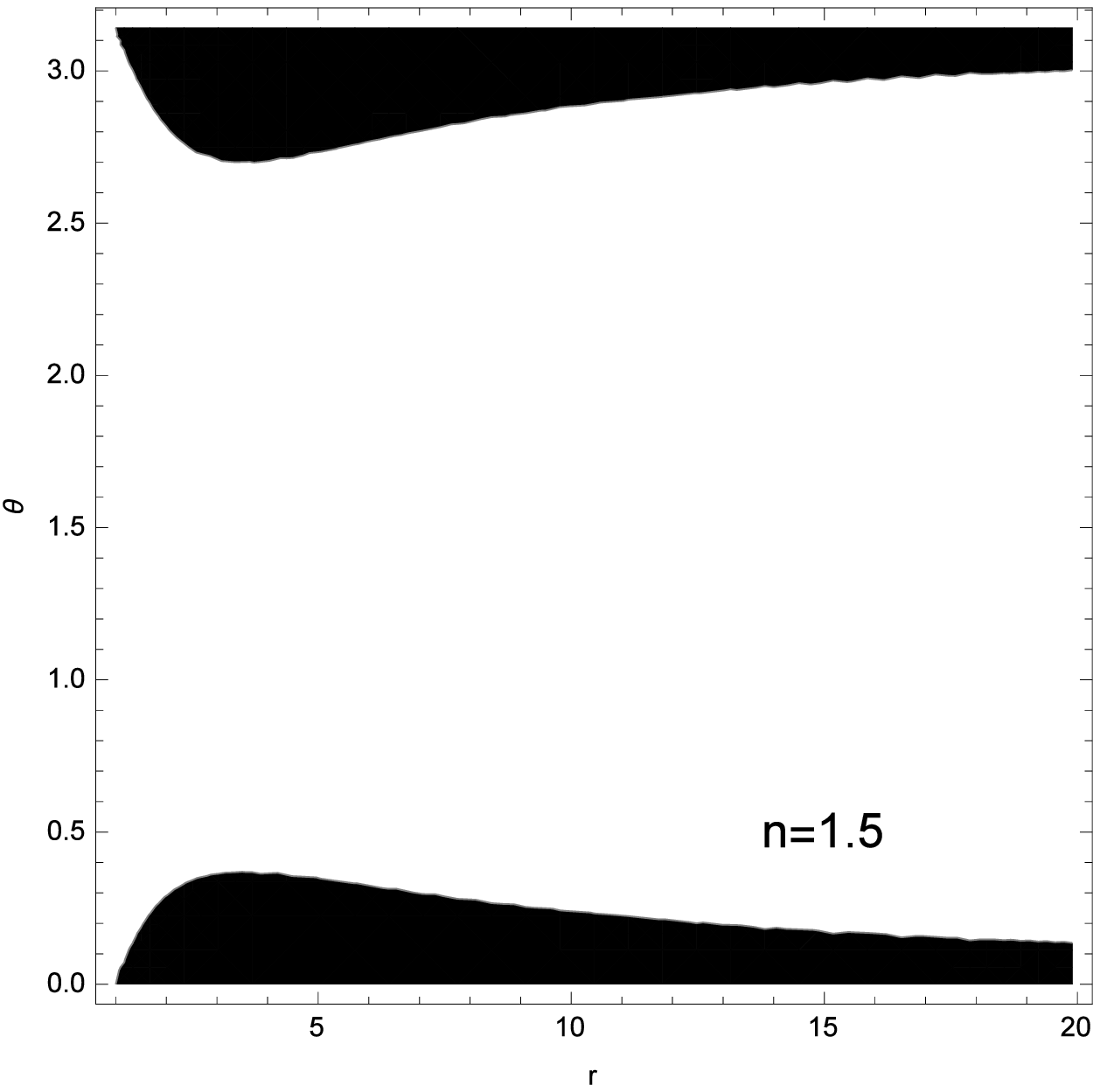}
\caption{The contour curves $g_{\phi\phi}=0$ for the extremal KTN black hole with different $n$. In the black region, the function $g_{\phi\phi}<0$ and then the closed timelike curves exist. In the white region, there is no closed timelike curves since $g_{\phi\phi}>0$. Here, we set $M=1$.}
\label{fig0}
\end{figure}
Due to the periodicity of angular coordinate, an azimuthal curve $\gamma=\{t=const$, $r=const$, $\theta=const\}$ is a closed curve with invariant length $ds^2=g_{\phi\phi}(2\pi)^2$. If the function $g_{\phi\phi}$ is negative, the integral curve with $(t, r, \theta)$ fixed becomes a closed timelike curve. For an extremal KTN black hole, we have
\begin{eqnarray}
g_{\phi\phi}=\frac{(r^2+M^2+2\hat{n}^2)^2\sin^2\theta-(r-M)^2(\sqrt{M^2+\hat{n}^2}\sin^2\theta-2\hat{n}\cos\theta)^2}
{r^2+(\hat{n}+\sqrt{M^2+\hat{n}^2}\cos\theta)^2}.
\end{eqnarray}
The sign of $g_{\phi\phi}$ depends on the parameters $M$, $\hat{n}$ and the coordinates $r$, $\theta$. In the equatorial plane, we find that $g_{\phi\phi}=\frac{r^4+(M^2+3\hat{n}^2)r^2+2M(M^2+\hat{n}^2)r+\hat{n}^2(3M^2+4\hat{n}^2)}
{r^2+\hat{n}^2}$, which always is positive. Thus, there is no such closed timelike curves in the equatorial plane for KTN black hole. In Fig.(\ref{fig0}), we present the regions in which $g_{\phi\phi}$ is negative or positive outside the event horizon of KTN black hole. In the black regions, the function $g_{\phi\phi}<0$ and then the closed timelike curves exist, but in the white region, $g_{\phi\phi}>0$ and there is no closed timelike curves. It is shown that the closed timelike curves disappear near the equatorial plane for different $n$. In this paper, we suppose a thin stationary and axisymmetric accretion disk around the a KTN black hole in the equatorial plane and focus on the case where the observer located in the position with a small deviation from the equatorial plane, which ensures that there is no closed timelike curves in the regions of the particle motion and the analysis method in Kerr case is still valid in these regions in KTN background. Moreover, this choice also avoid that a particle moving along geodesic does pass through  the conical singularities located in the axis of symmetry.

In the KTN black hole spacetime, the timelike geodesic equation for  particle in the spacetime (\ref{metric1}) can be expressed as
\begin{eqnarray}
&&\dot{t}=\frac{g_{t\phi}\hat{L}+g_{\phi\phi}\hat{E}}{g^2_{t\phi}-g_{tt}g_{\phi\phi}},\nonumber\\
&&\dot{\phi}=-\frac{g_{t\phi}\hat{E}+g_{tt}\hat{L}}{g^2_{t\phi}-g_{tt}g_{\phi\phi}},\nonumber\\
&&g_{\hat{r}\hat{r}}\dot{\hat{r}}^2+g_{\theta\theta}\dot{\theta}^2=V_{eff}(\hat{r},\theta)=
\frac{g_{\phi\phi}\hat{E}^2+2g_{t\phi}\hat{E}\hat{L}+g_{tt}\hat{L}^2}{g^2_{t\phi}-g_{tt}g_{\phi\phi}}-\mu^2,
\end{eqnarray}
where $\hat{E}$ and $\hat{L}$, respectively, are the conserved energy  and the conserved $z$-component of the  angular momentum.  The overhead dot represents a derivative with respect to the affine parameter.
A circular orbit for a particle in the equatorial plane $\theta=\frac{\pi}{2}$ must satisfy
\begin{eqnarray}
V_{eff}(\hat{r},\theta)\bigg|_{\theta=\frac{\pi}{2}}=0,\;\;\;\;\;\;\;\;\;\;\;\;\;\;\;\;
\frac{dV_{eff}(\hat{r},\theta)}{d\hat{r}}\bigg|_{\theta=\frac{\pi}{2}}=0.
\end{eqnarray}
In the KTN spacetime, the four-velocity of particle moving along a circular orbit with a radius $\hat{r}=\hat{r}_s$ in the equatorial plane  can be given by \cite{P27}
\begin{eqnarray}
u_s&=&u^t_s(\partial_t+\Omega_s\partial_\phi), \nonumber\\
u^t_s&=&\frac{\sqrt{\hat{r}}(\hat{r}^2+\hat{n}^2)+a\sqrt{W}}
{\sqrt{(\hat{r}^2+\hat{n}^2)[\hat{r}^3-3M\hat{r}^2-\hat{n}^2(3\hat{r}-M)+2a\sqrt{\hat{r}W}]}},\nonumber\\
\Omega_s&=&\frac{\sqrt{W}}{(\hat{r}^2+\hat{n}^2)\sqrt{\hat{r}}+a\sqrt{W}},
\end{eqnarray}
with
\begin{eqnarray}
W&=&(\hat{r}^2-\hat{n}^2)M+2\hat{n}^2\hat{r},
\end{eqnarray}
where the subscript $s$  denotes ``source". Together with the condition  $\frac{d^2V_{eff}(\hat{r},\theta)}{d\hat{r}^2}|_{\theta=\frac{\pi}{2}}=0$, one can get the radius of the innermost stable circular orbit $\hat{r}_{ISCO}$ for the particle around the black hole, which can not be analytically given in the KTN spacetime (\ref{metric1}). The circular orbit with radius  $\hat{r}_s$ only larger than $\hat{r}_{ISCO}$ is stable for the timelike particle.

As in Ref.\cite{P19}, we here assume that all of electromagnetic radiation caused by the source propagate along null geodesic from the disk to the distant observer located in the position with the polar coordinate $r_0$ and the polar angle $\theta_0$. There are also three conserved quantities for these photons from the source, i.e., the energy $E$, the angular momentum $L$ and the Cater constant
\begin{eqnarray}
Q=p^2_{\theta}-\cos^2{\theta}\bigg[(a^2-4\hat{n}^2\csc^2{\theta})p^2_t-\csc^2{\theta}p^2_{\phi}\bigg]
-4\hat{n}\cos{\theta}p_t(ap_t-\csc^2{\theta}p_{\phi}).
\end{eqnarray}
With two rescaled quantities,
\begin{eqnarray}
\hat{\lambda}&=&\frac{L}{E},\quad \quad\quad \hat{q}=\frac{\sqrt{Q}}{E},
\end{eqnarray}
the null geodesic equation for the photon  can be further rewritten as
\begin{eqnarray}
\int^{\hat{r}_0}_{\hat{r}_s}\frac{d\hat{r}}{\pm\sqrt{\hat{R}(\hat{r})}}=
\int^{\theta_0}_{\theta_s}\frac{d\theta}{\pm\sqrt{\hat{\Theta}(\theta)}},
\label{ced}
\end{eqnarray}
with
\begin{eqnarray}
\hat{R}(\hat{r})&=&(\hat{r}^2+a^2+\hat{n}^2-a\hat{\lambda})^2-(\hat{r}^2-2M\hat{r}+a^2
-\hat{n}^2)[\hat{q}^2+(\hat{\lambda}-a)^2],\\
\hat{\Theta}(\theta)&=&\hat{q}^2+\cos^2{\theta}\bigg(a^2-\frac{4\hat{n}^2+\hat{\lambda}^2}{\sin^2{\theta}}\bigg)
+4\hat{n}\cos{\theta}\bigg(a-\frac{\hat{\lambda}}{\sin^2{\theta}}\bigg).\label{thetaq}
\end{eqnarray}
Making use of the energy of  photons at the source $E_s=-p\cdot u_s$, and the energy at the distant observer $E_o=E=-p_t$, one can define the redshift factor $g$ as \cite{P28}
\begin{eqnarray}
g&\equiv&\frac{E_o}{E_s}=\frac{\sqrt{(\hat{r}^2+\hat{n}^2)
[\hat{r}^3-3M\hat{r}^2-\hat{n}^2(3\hat{r}-M)+2a\sqrt{\hat{r}W}]}}{
\sqrt{\hat{r}}(\hat{r}^2+\hat{n}^2)+(a-\hat{\lambda}) \sqrt{W}}.
\label{gred}
\end{eqnarray}
Like in Ref.\cite{P19}, the conserved quantities  $\hat{\lambda}$ and $\hat{q}$ can be written mathematically as  functions of the radius $r_s$ and the redshift factor $g$.
Moreover, it is obvious that the photons which can arrive at observer determine an image of the emitter on the observer's screen. The position of the image on the observer's local sky can be described by the angular coordinates $(\alpha,\beta)$, which are related to $(\hat{\lambda},\hat{q})$ by \cite{KTNapp8} (also see in the Appendix)
\begin{eqnarray}
\alpha&=&-\frac{\hat{\lambda}}{\sin{\theta_o}},\quad\quad\quad \beta=\pm\sqrt{\hat{\Theta}(\theta_o)}.
\label{tianq}
\end{eqnarray}
Thus, the solid angle for a light ray from the disk can be expressed as
\begin{eqnarray}
d\Omega&=&\frac{1}{\hat{r}^2_o}d\alpha d\beta=\frac{1}{\hat{r}^2_o}
\bigg{\arrowvert}\frac{\partial(\alpha,\beta)}{\partial(\hat{\lambda},\hat{q})}\bigg{\arrowvert}
\bigg{\arrowvert}\frac{\partial(\hat{\lambda},\hat{q})}{\partial(\hat{r}_s,g)}\bigg{\arrowvert}  d\hat{r}_s dg=\frac{\hat{q}}{|\beta|\hat{r}^2_o\sin{\theta}_o}\bigg{\arrowvert}
\frac{\partial(\hat{\lambda},\hat{q})}{\partial(\hat{r}_s,g)}\bigg{\arrowvert}d\hat{r}_s dg.
\end{eqnarray}
The specific flux of the ray of light measured at observer is
\begin{eqnarray}
dF_o=I_od\Omega=g^3I_sd\Omega,
\end{eqnarray}
where the relationship between the surface brightness seen by the distant observer $I_o$ and  the surface brightness evaluated at the source $I_s$ is obtained from Liouville's theorem on the invariance of the phase space density of photons. For the simplicity, as in Ref. \cite{P19}, we can suppose that the disk's specific intensity is monochromatic at energy $E_\star$ and isotropic with surface emissivity $\mathcal{E}(\hat{r}_s)$, i.e.,
\begin{eqnarray}
I_s&=&\mathcal{E}(\hat{r}_s)\delta(E_s-E_\star)=g\mathcal{E}(E_o-gE_\star),
\end{eqnarray}
and then obtain the flux at observer is
\begin{eqnarray}
F_o&=&\frac{g^4}{\hat{r}^2_o\sin{\theta}_o}\int\frac{\hat{q}}{|\beta|}
\bigg{\arrowvert}\frac{\partial(\hat{\lambda},\hat{q})}{\partial(\hat{r}_s,g)}\bigg{\arrowvert}
\mathcal{E}(\hat{r}_s)d\hat{r}_s,
\end{eqnarray}
where a factor of $E_\star$ is absorbed into $\mathcal{E}(\hat{r}_s)$. The integral is to be calculated over the radial extent of the accretion disk starting from the innermost stable circular orbit. Obviously, the flux $F_o$ depends on the radius of circular orbit of ``source" $ \hat{r}_s$, the parameters of photon $\hat{q}$ and $\hat{\lambda}$, the position of observer $ (\hat{r}_o, \theta_o)$ and  the disk model described by function $\mathcal{E}(\hat{r}_s)$.

Let us now focus on the case of extremal KTN black hole and then probe the emissions
originating from the innermost part of the accretion disk near the innermost stable circular orbit. In this special case, one can get an analytic expression for the Jacobian determinant $|\partial(\hat{\lambda},\hat{q})/\partial(\hat{r_s},g)|$, which is convenient to study analytically the properties of the flux $F_o$. For the extremal KTN black hole with $a=\sqrt{M^2+\hat{n}^2}$, one can introduce dimensionless radial coordinate and parameters \cite{P15,P17,P19}
\begin{eqnarray}
r&=&\frac{\hat{r}-M}{M},\quad \quad\lambda=1-\frac{\hat{\lambda}}{2M\sqrt{1+n^2}},\quad \quad q^2=3-n^2-\frac{\hat{q}^2}{M^2},
\end{eqnarray}
where $n\equiv\frac{\hat{n}}{M}$.
With these quantities,  the geodesic equation (\ref{ced}) becomes
\begin{eqnarray}
\int^{r_0}_{r_s}\frac{dr}{\pm\sqrt{R(r)}}=
\int^{\theta_0}_{\theta_s}\frac{d\theta}{\pm\sqrt{\Theta(\theta)}},
\end{eqnarray}
 with
\begin{eqnarray}\label{Rsb}
R(r)&=&r^4+4r^3+\bigg[q^2+4\lambda(2-\lambda)(n^2+1)
\bigg]r^2
+8\lambda r(n^2+1)+4\lambda^2(n^2+1)^2,\\
\Theta(\theta)&=&3-q^2+\cos^2{\theta}-n^2\sin^2{\theta}-
4\bigg[(\lambda-1)^2+n^2(\lambda^2-2\lambda+2)\bigg]\cot^2{\theta}\nonumber\\
&&+4n\sqrt{1+n^2}\bigg[2(\lambda-1)+\sin^2{\theta}\bigg]
\frac{\cos{\theta}}{\sin^2{\theta}}.
\end{eqnarray}
As done in Refs. \cite{P17,P19}, for the electromagnetic signal from the near-horizon region of extremal KTN black hole, one can obtain that $\hat{\lambda}=2M\sqrt{1+n^2}+\mathcal{O}$ , where $\mathcal{O}$ is a small quantity. Therefore, when the source point is near the horizon ($r_s\ll 1$) and the observation point $r_o$ is in the spatial infinite far region ($r_o\gg1$), from Eq. (\ref{tianq}), we can find  that the electromagnetic signal from the near-horizon region lies in the NHEKTN line on the observer's sky with the coordinates
\begin{eqnarray}
\alpha&=&-2M\sqrt{n^2+1}\csc{\theta_o},\nonumber\\
\beta&=&\pm M\bigg[3-q^2+\cos^2{\theta_o}-n^2\sin^2{\theta_o}-
4(2n^2+1)\cot^2{\theta_o}-4n\sqrt{n^2+1}
\frac{(2-\sin^2{\theta_o})\cos{\theta_o}}{\sin^2{\theta_o}}\bigg]^{1/2}.\label{betas}
\end{eqnarray}
In the Kerr black hole spacetime, for a fixed redshift factor $g$, the dominant contribution in the flux of electromagnetic emission from sources at $r_s$  comes from the photons whose $r_s$ is a near-region
radial turning point for the null geodesics \cite{P17}. Here, we assume that it is still valid in the case of the KTN black hole. In the near-horizon region, $R(r)$ (\ref{Rsb}) can be approximated as
\begin{eqnarray}
R_n(r)&=&q^2r^2+8\lambda(n^2+1) r+4\lambda^2(n^2+1)^2,
\end{eqnarray}
and then the turning point $r_s$ in radial direction is obtained by solving $R(r)=0$,
\begin{eqnarray}
r_s&=&-\frac{2\lambda}{q^2}(n^2+1)(2+\sqrt{4-q^2}).\label{rrs}
\end{eqnarray}
Moreover, from Eq. (\ref{gred}) with the condition $r_s\ll1$, we have
\begin{eqnarray}
\lambda&=&\sqrt{\frac{3-n^2}{n^2+1}}\bigg(\frac{1}{g}-\sqrt{\frac{3-n^2}{n^2+1}}\bigg)
\,\frac{r_s}{4}.\label{rrs2}
\end{eqnarray}
Combining equation Eq. (\ref{rrs})-(\ref{rrs2}), we obtain
\begin{eqnarray}
q&=&\frac{\sqrt{3-n^2}}{2g}\bigg[\bigg(g-\sqrt{\frac{n^2+1}{3-n^2}}\bigg)
\bigg((n^2+5)g+\sqrt{3-n^2}\sqrt{n^2+1}\bigg)\bigg]^{\frac{1}{2}},\label{qs3}
\end{eqnarray}
which $q$ depends on both $g$ and the NUT charge $n$. As $n$ disappears, we find that $q$ is only a function of $g$, which is consistent with that obtained in Ref.\cite{P19}. From Eqs.(\ref{betas}), (\ref{rrs}) and (\ref{qs3}) , we find that $\lambda<0$, and
\begin{eqnarray}
&&0<q<q_c=\bigg[3+\cos^2{\theta_o}-n^2\sin^2{\theta_o}-
4(2n^2+1)\cot^2{\theta_o}-
\frac{4n\sqrt{n^2+1}(2-\sin^2{\theta_o})\cos{\theta_o}}{\sin^2{\theta_o}}\bigg]^{1/2},\label{qcc}\\
&&\sqrt{\frac{n^2+1}{3-n^2}}<g<\frac{\sqrt{(n^2+1)(3-n^2)}(n^2+1+2\sqrt{4-q^2_{\text{max}}})}{
(3-n^2)(n^2+5)+4q^2_{\text{max}}}\label{gcc}.
\end{eqnarray}
Here $q_{\text{max}}$ is the maximum value of $q_c$ in Eq. (\ref{qcc}), which has no analytical expression in the KTN spacetime. As the NUT charge $n\rightarrow 0$, one can find that $q_{\text{max}}=\sqrt{3}$, which reduces to that in Kerr case.
Inserting Eq.(\ref{qs3}) into Eq.(\ref{betas}), we obtain the angular coordinate $\beta$ of the points in the NHEKTN line on the observer's sky
\begin{eqnarray}
\beta&=&\pm M\bigg[3+\cos^2{\theta_o}-n^2\sin^2{\theta_o}-
4(2n^2+1)\cot^2{\theta_o}-4n\sqrt{n^2+1}
\frac{(2-\sin^2{\theta_o})\cos{\theta_o}}{\sin^2{\theta_o}}\nonumber\\
&-&\frac{3-n^2}{4g^2}\bigg(g-\sqrt{\frac{n^2+1}{3-n^2}}\bigg)
\bigg((n^2+5)g+\sqrt{3-n^2}\sqrt{n^2+1}\bigg)\bigg]^{1/2}.
\end{eqnarray}
From Eqs. (\ref{rrs2})-(\ref{qs3}), we find that the second Jacobian can be expressed as
\begin{eqnarray}
\frac{\partial(\lambda,q)}{\partial(r_s,g)}&=&\frac{(3-n^2)}{16qg^4}
\bigg(\sqrt{3-n^2}+\sqrt{n^2+1}g\bigg)\bigg(\sqrt{n^2+1}-\sqrt{(3-n^2)}g\bigg),
\end{eqnarray}
and then the flux at the observer can be further rewritten as
\begin{eqnarray}
F_o&=&\frac{(3-n^2)\sqrt{n^2+1}}{8r^2_o\sin{\theta_o}|\beta|}\bigg(\sqrt{3-n^2}+\sqrt{n^2+1}g\bigg)
\bigg(\sqrt{3-n^2}g-\sqrt{n^2+1}\bigg)\int\mathcal{E}(r_s)dr_s.
\end{eqnarray}
The dependence of $F_o$ on $g$ implies that the flux at different points on the observer's  NHEKTN line is also determined by photons of different energy. In terms of  the discussions in Ref. \cite{P19}, the role of the integral $\int\mathcal{E}(r_s)dr_s$ in the above equation is only to fix the overall scale of $F_o$ for the particular disk model by the surface emissivity function $\mathcal{E}$. This integral is divergent for the extremal black hole, which are likely caused by
the caustics \cite{P17} and could be regulated by diffraction effects in wave optics. The factor before the integral $\int\mathcal{E}(r_s)dr_s$ is independent of disk models and presents the universal features of electromagnetic line emissions from the near-horizon region of extremal KTN black hole. Thus,
the universal properties of the observed flux $F_o$ can be described by
\begin{eqnarray}
F_o&\varpropto&\frac{(3-n^2)\sqrt{n^2+1}}{8r^2_o\sin{\theta_o}|\beta|}\bigg(\sqrt{3-n^2}+\sqrt{n^2+1}g\bigg)
\bigg(\sqrt{3-n^2}g-\sqrt{n^2+1}\bigg).
\end{eqnarray}
Obviously, $F_o$ is a function of both the redshift factor $g$ and the angular coordinate $\beta$ on
the NHEKTN line, which depend on the NUT charge $n$ and the observer's angular position $\theta_o$. The dependence of the flux $F_o$ on the $n$  could provide a potential observable to examine whether the real spacetime belongs to such kind of KTN spacetime.

\section{Effects of NUT charge on the electromagnetic line emissions from near-horizon region}

 We are now in position to study the effects of the NUT charge on electromagnetic line emissions from near-horizon region in the extremal KTN black hole spacetime.
\begin{figure}
\center
\includegraphics[width=5cm]{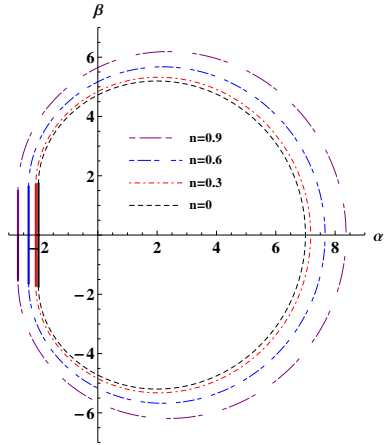}\;\;\;\;
\caption{ The NHEKTN line in the KTN black hole shadow for different NUT charge $n$. Here, we set $M=1$.}
\label{fig01}
\end{figure}
\begin{figure}
\center
\subfigure[]{\includegraphics[width=4.5cm]{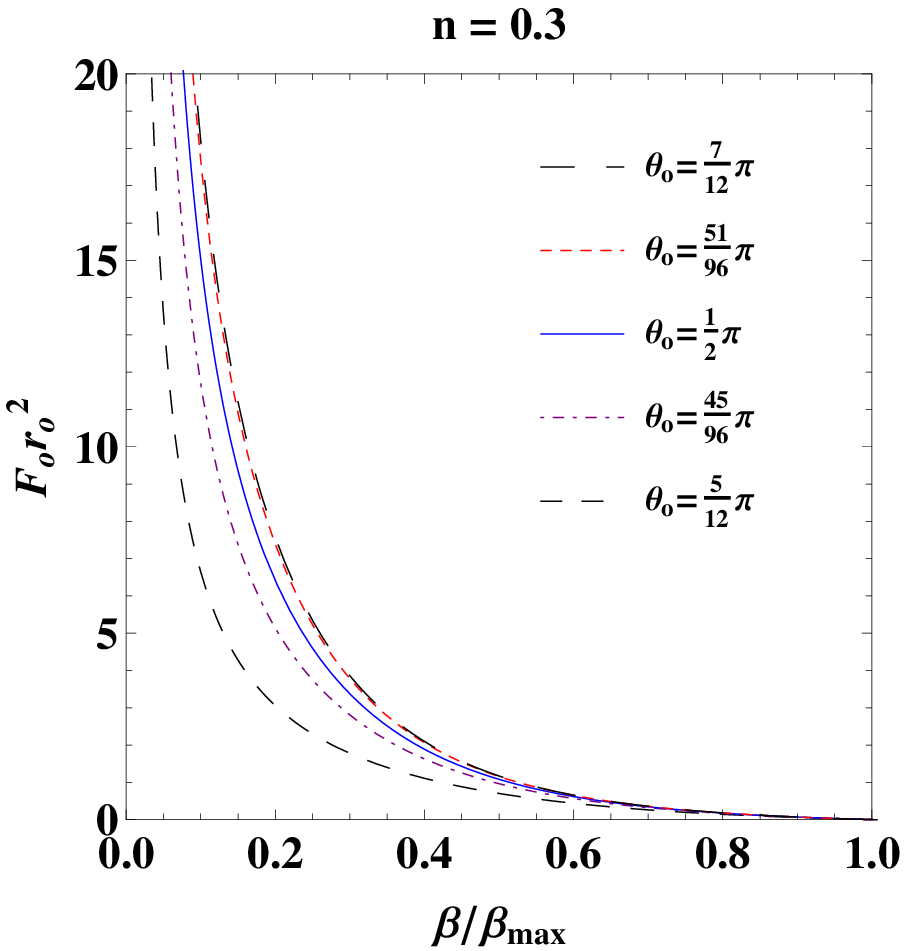}}
\;\;\;\;\subfigure[]{\includegraphics[width=4.5cm]{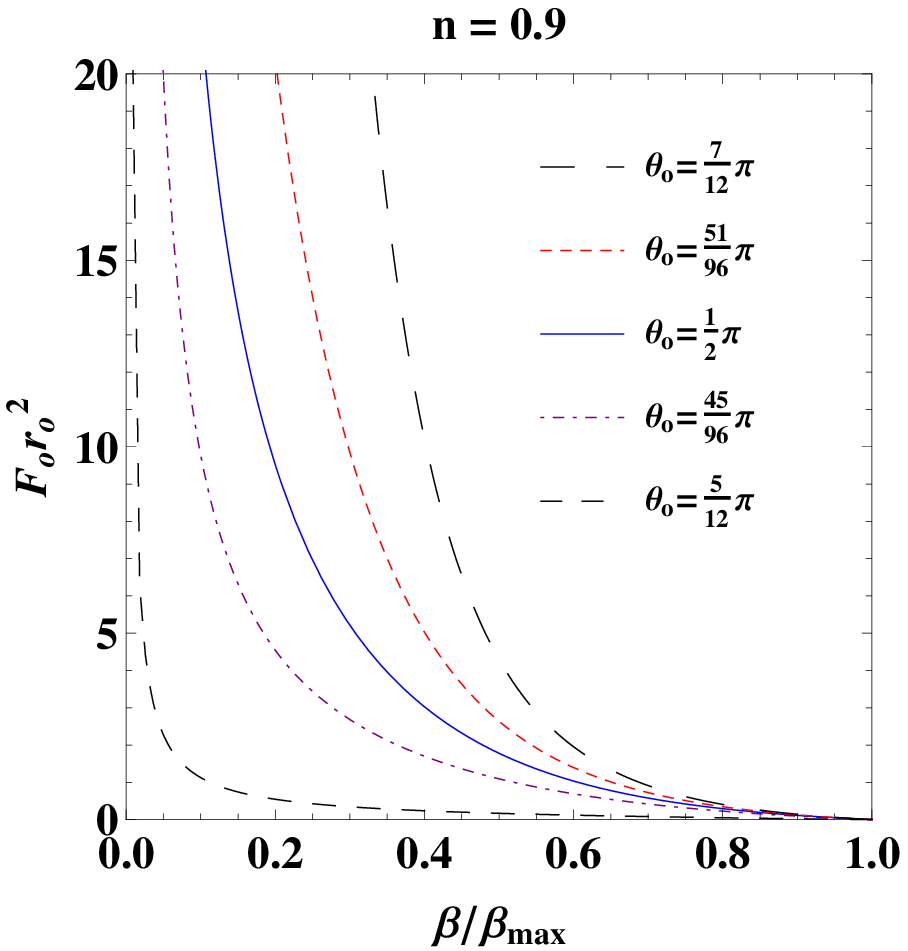}}
\subfigure[]{\includegraphics[width=4.5cm]{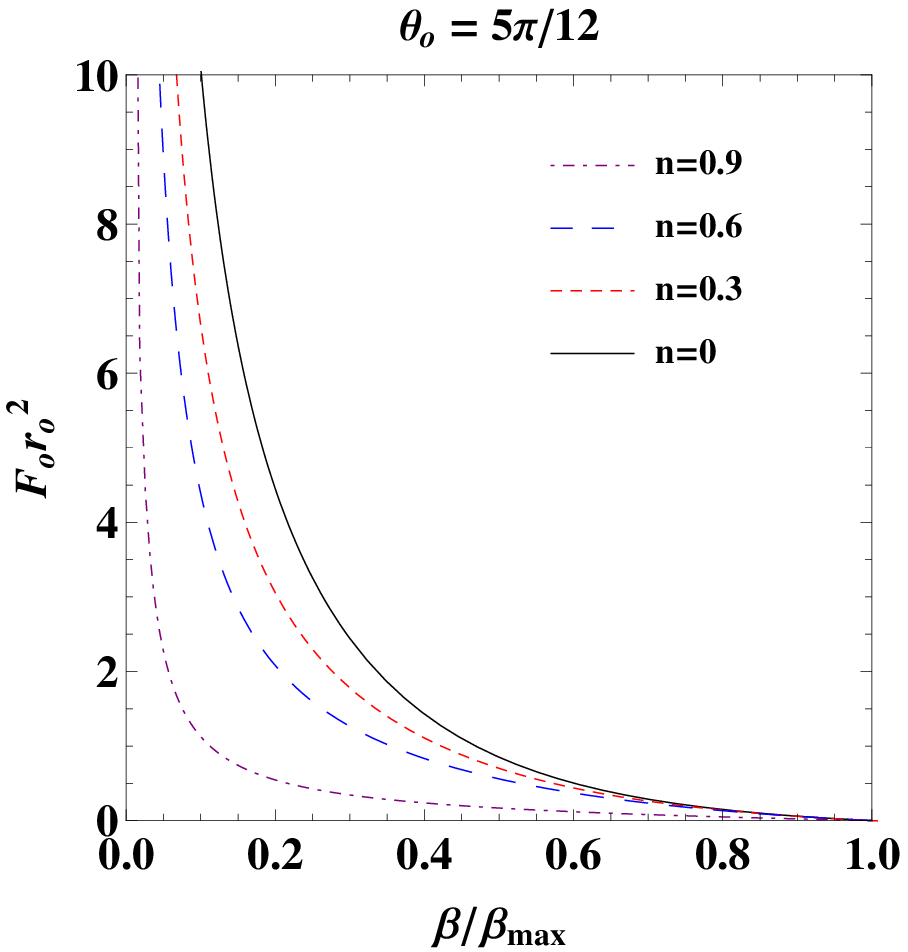}}
\\
\subfigure[]{\includegraphics[width=4.5cm]{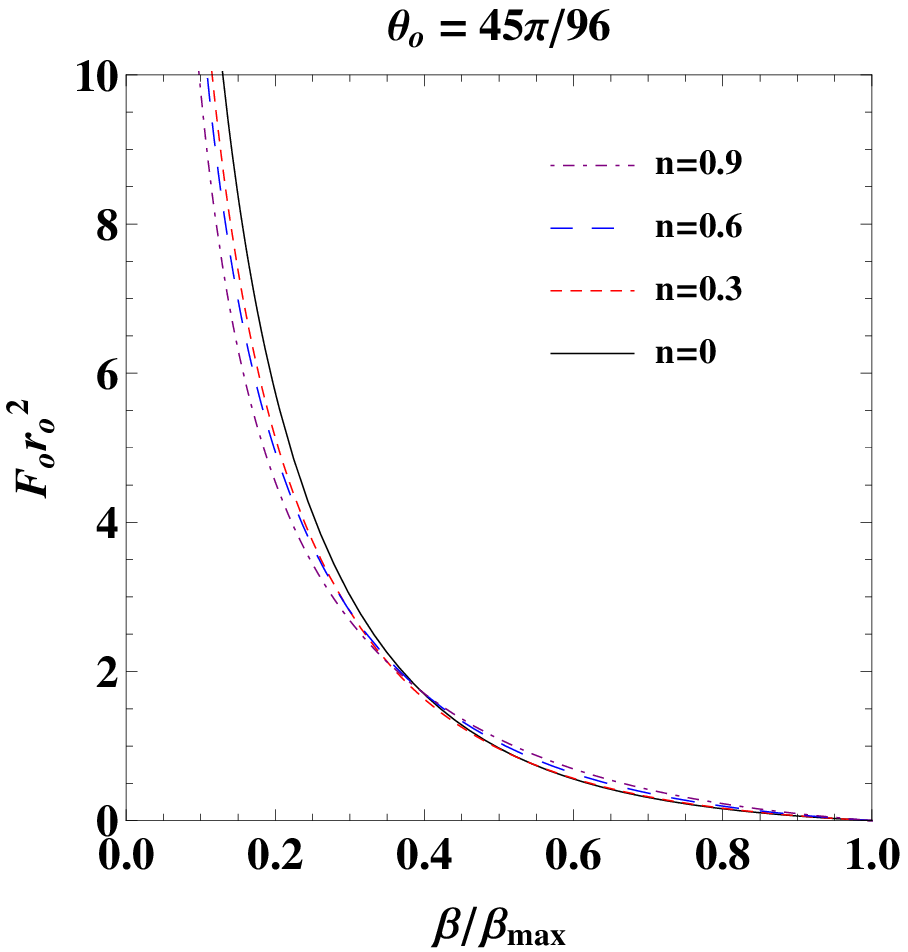}}
\;\;\;\;
\subfigure[]{\includegraphics[width=4.5cm]{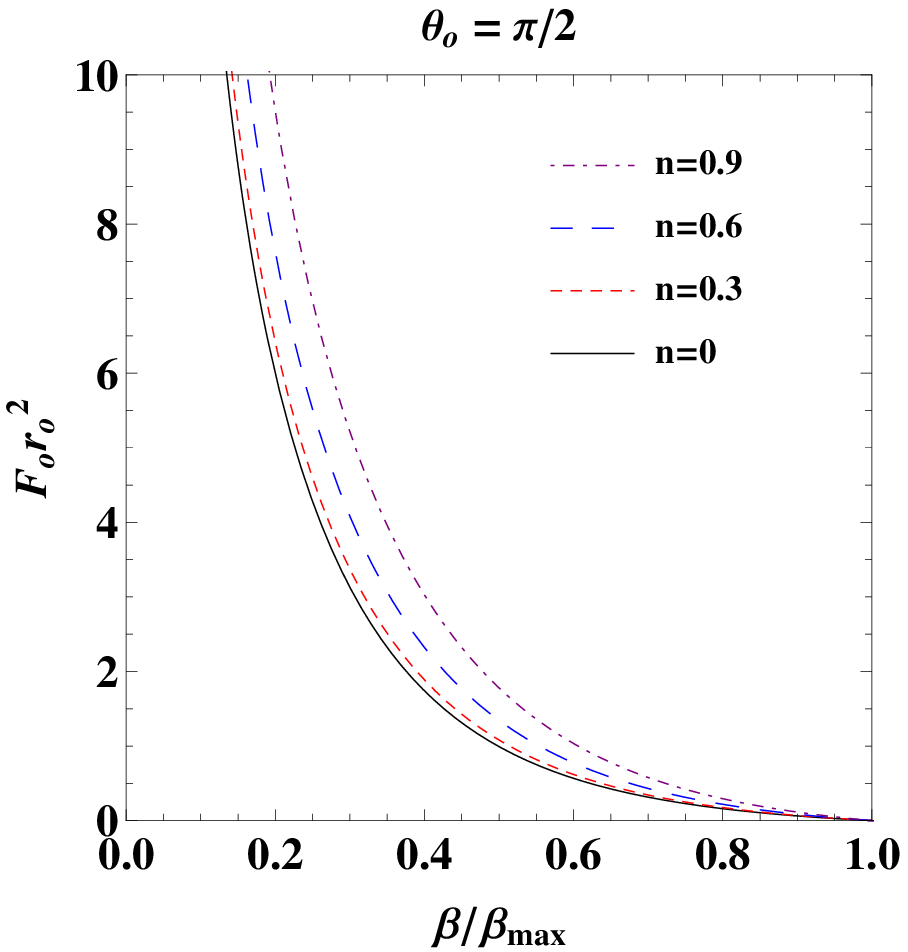}}
\subfigure[]{\includegraphics[width=4.5cm]{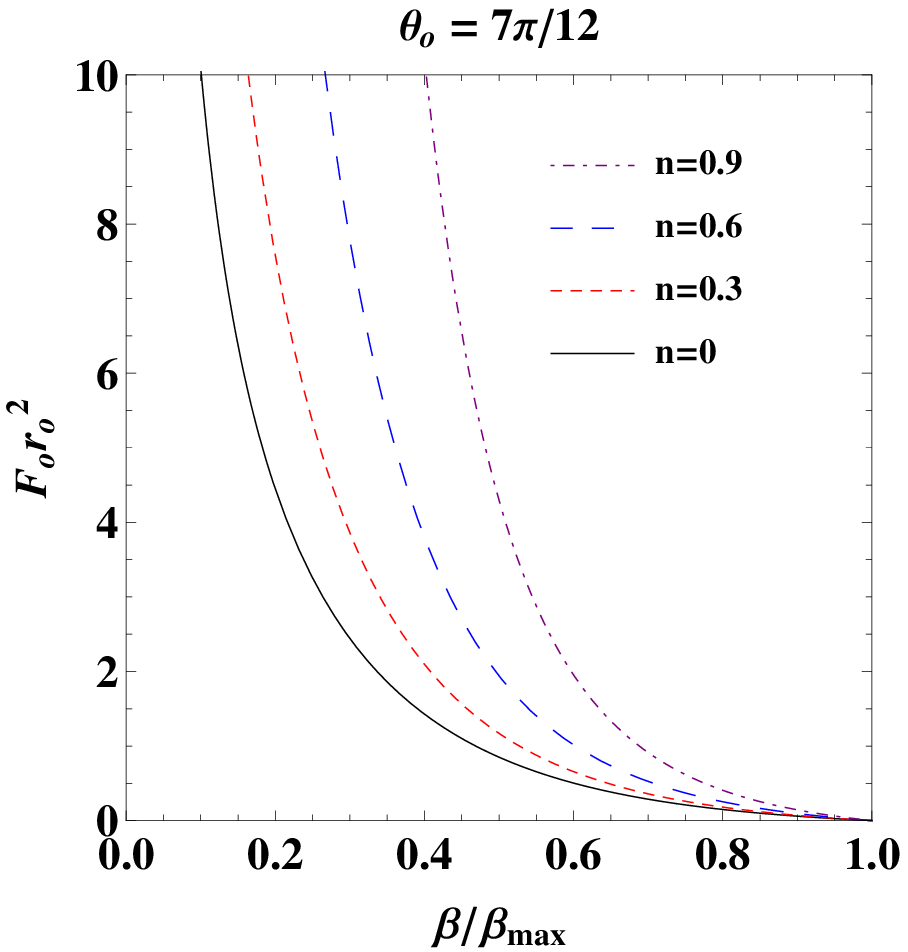}}
\caption{The change of the flux $F_o$ at observer with the angular coordinate $\beta/\beta_{max}$ for different NUT charge $n$ and the observer's angular position $\theta_o$ as the source lies in the NHEKTN line. Here, we set $M=1$.}
\label{fig1}
\end{figure}
In Fig. (\ref{fig1}), we present the change of the flux $F_o$ at observer with the angular coordinate ratio $\beta/\beta_{max}$ for different NUT charge $n$ and the observer's angular position $\theta_o$ as the source lies in the NHEKTN line, which is a vertical edge in the shadow depicted in solid line in Fig. (\ref{fig01}). $\beta_{max}$ is the maximum value of the coordinate $\beta$ of the image located in the NHEKTN line. From Fig. (\ref{fig1}), we find that the flux $F_o$ decreases with the ratio $\beta/\beta_{max}$, which means that the brightest electromagnetic line at observer is emitted from the sources located in the equatorial plane. Moreover, for the fixed NUT charge $n$, one can find from Figs. (\ref{fig1}a) and (\ref{fig1}b) that the flux $F_o$ increases with the angular coordinate $\theta_o$ of the observer's position, which implies that electromagnetic line emission for the observer in the Southern Hemisphere is brighter than that in the Northern one. This is different from that in the Kerr case where
the electromagnetic line emission is the brightest for the observer in the equatorial plane. The main reason is that the equatorial plane is no more a symmetry plane of the KTN spacetime due to the presence of the NUT charge. Theoretically, this new properties of electromagnetic line emission from the near-horizon region could help us to detect the gravitomagnetic monopole.
\begin{figure}
\center
\includegraphics[width=4cm]{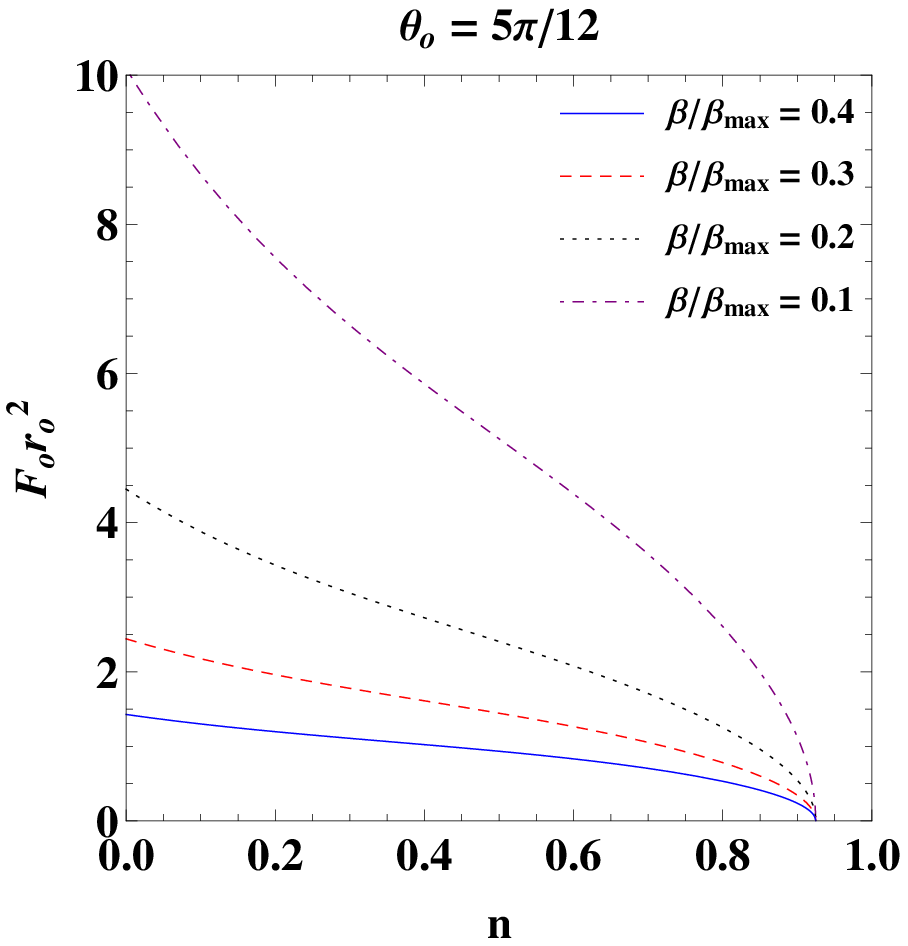}
\includegraphics[width=4.0cm]{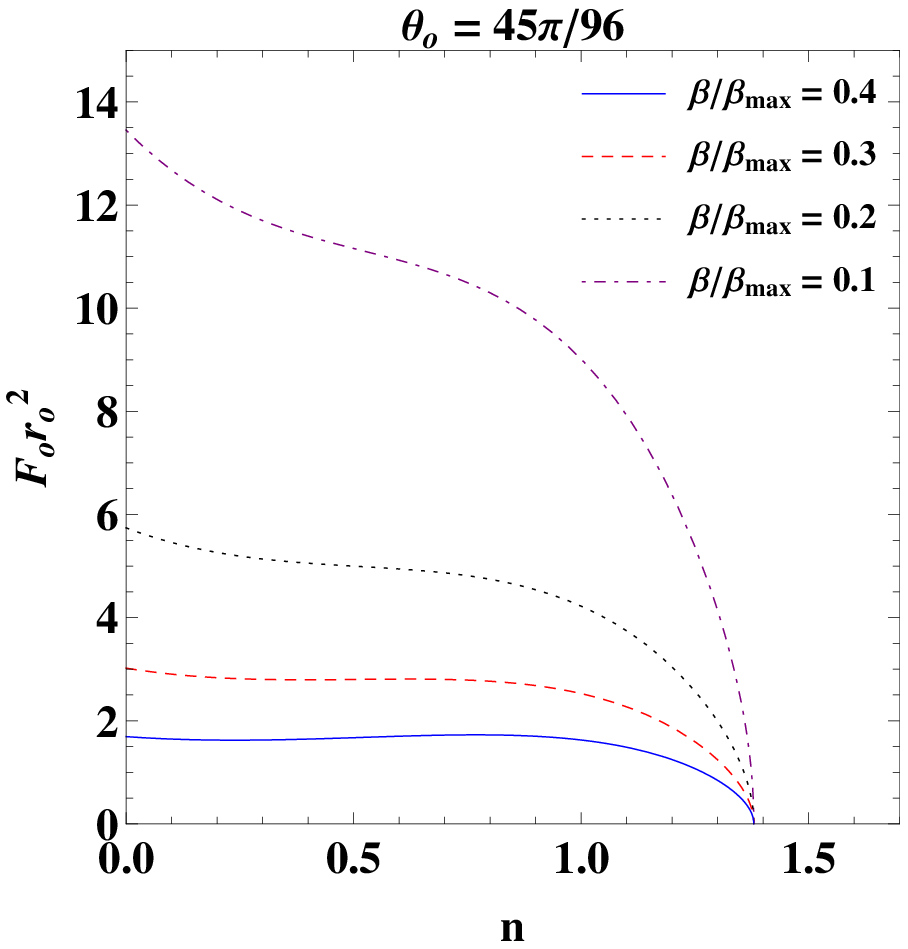}
\includegraphics[width=4.0cm]{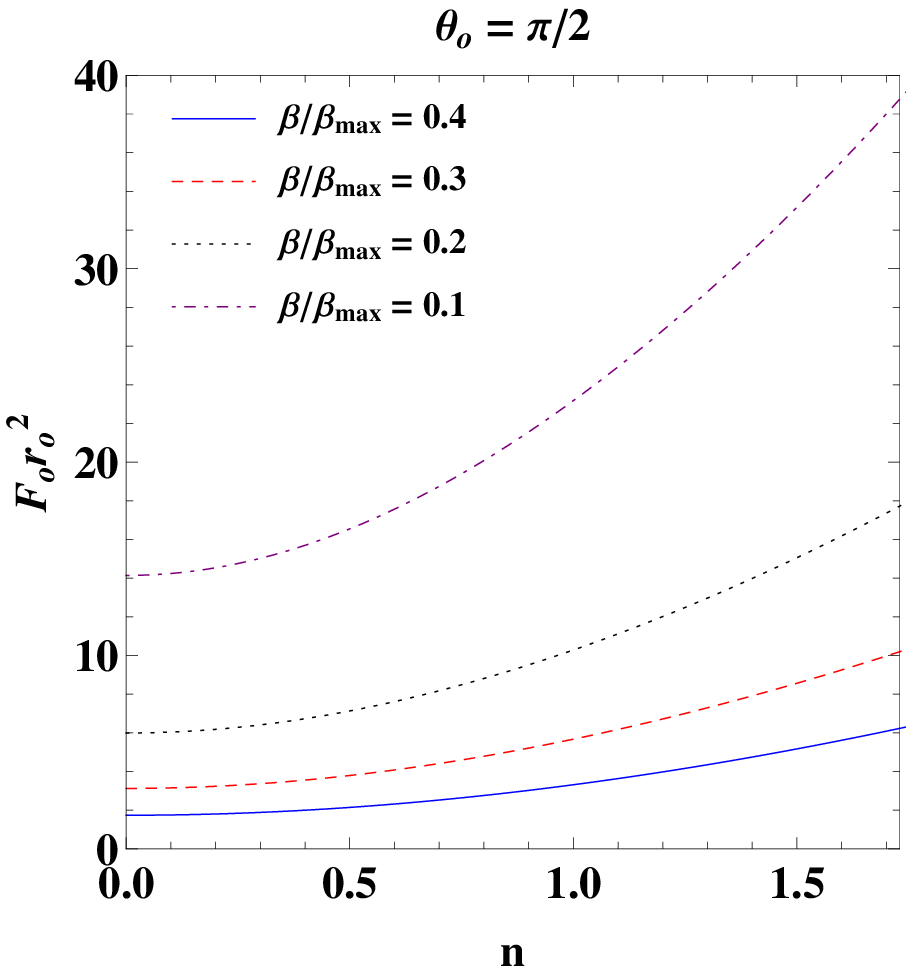}
\includegraphics[width=4cm]{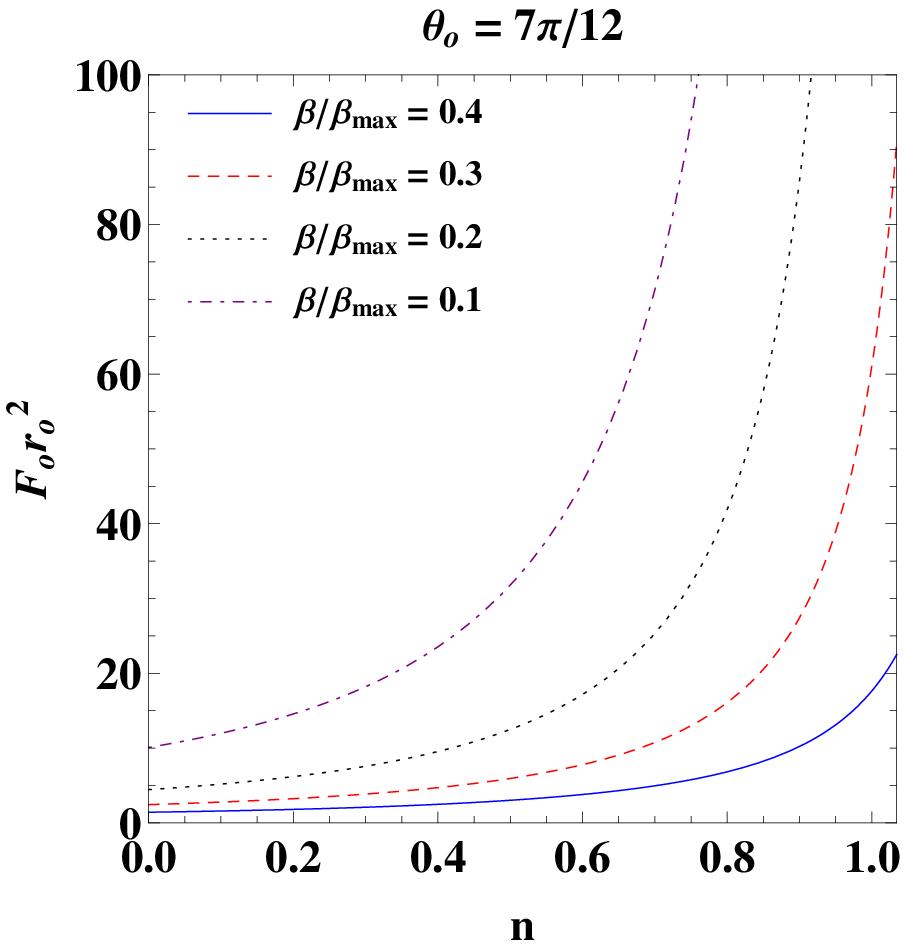}
\\
\includegraphics[width=4cm]{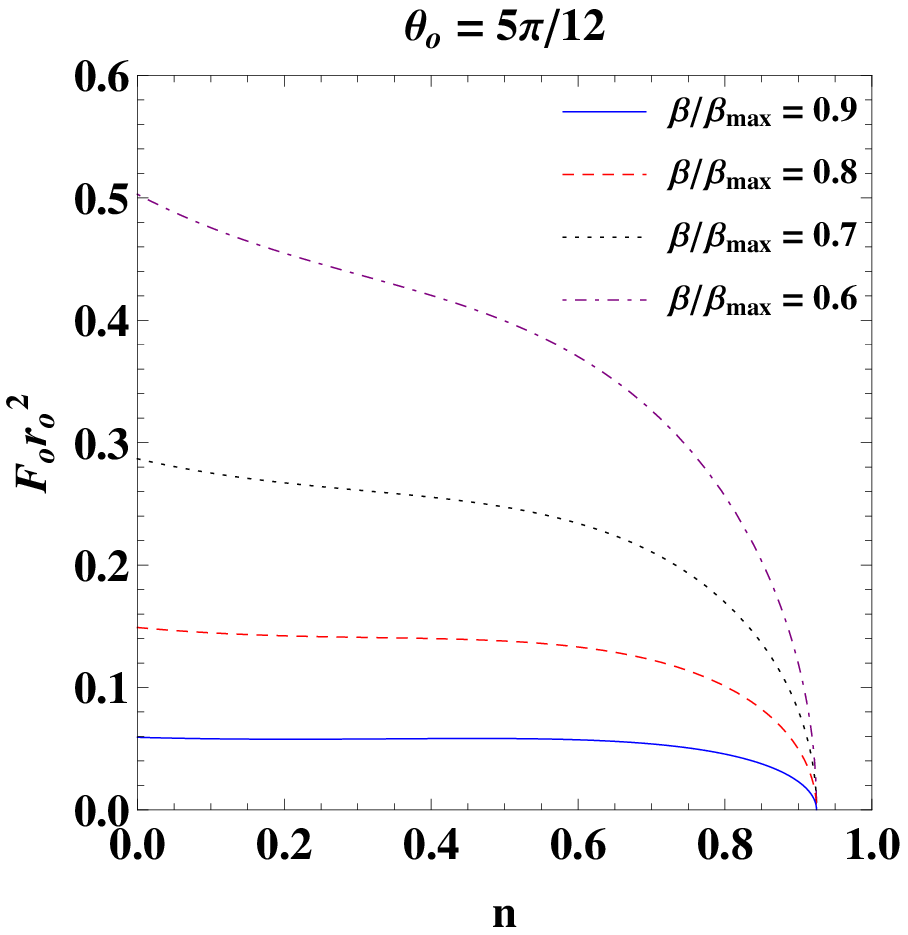}
\includegraphics[width=4.0cm]{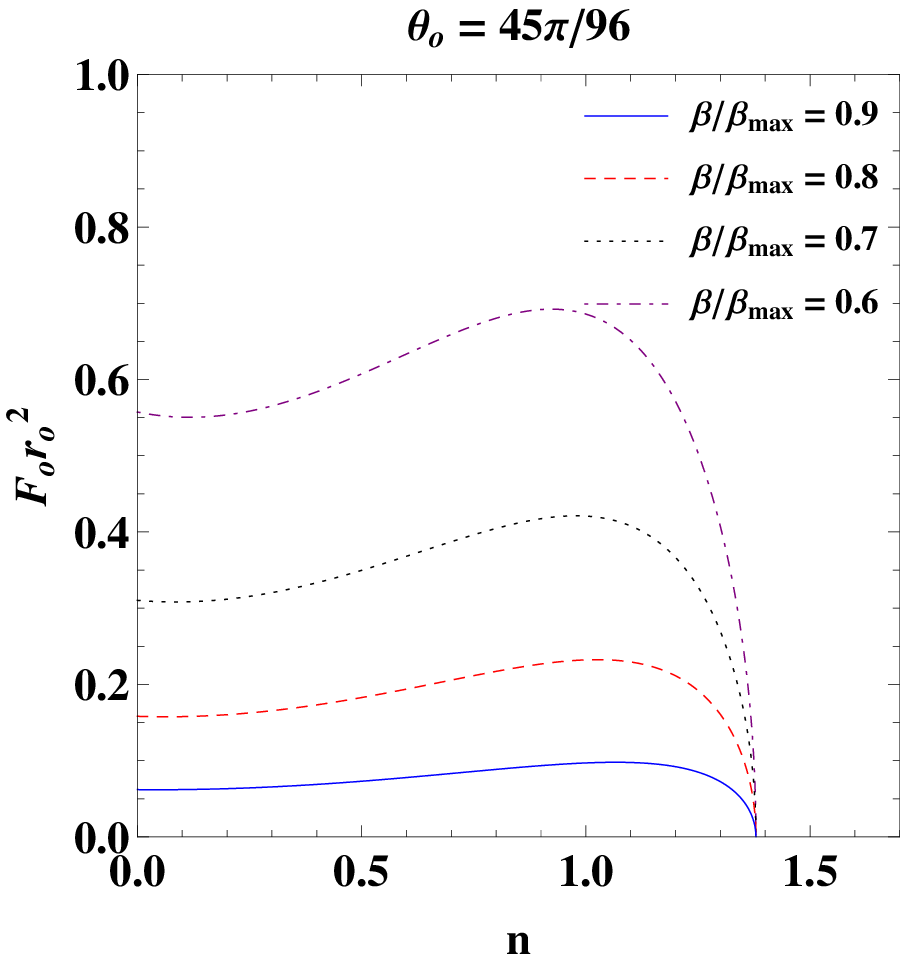}
\includegraphics[width=4.0cm]{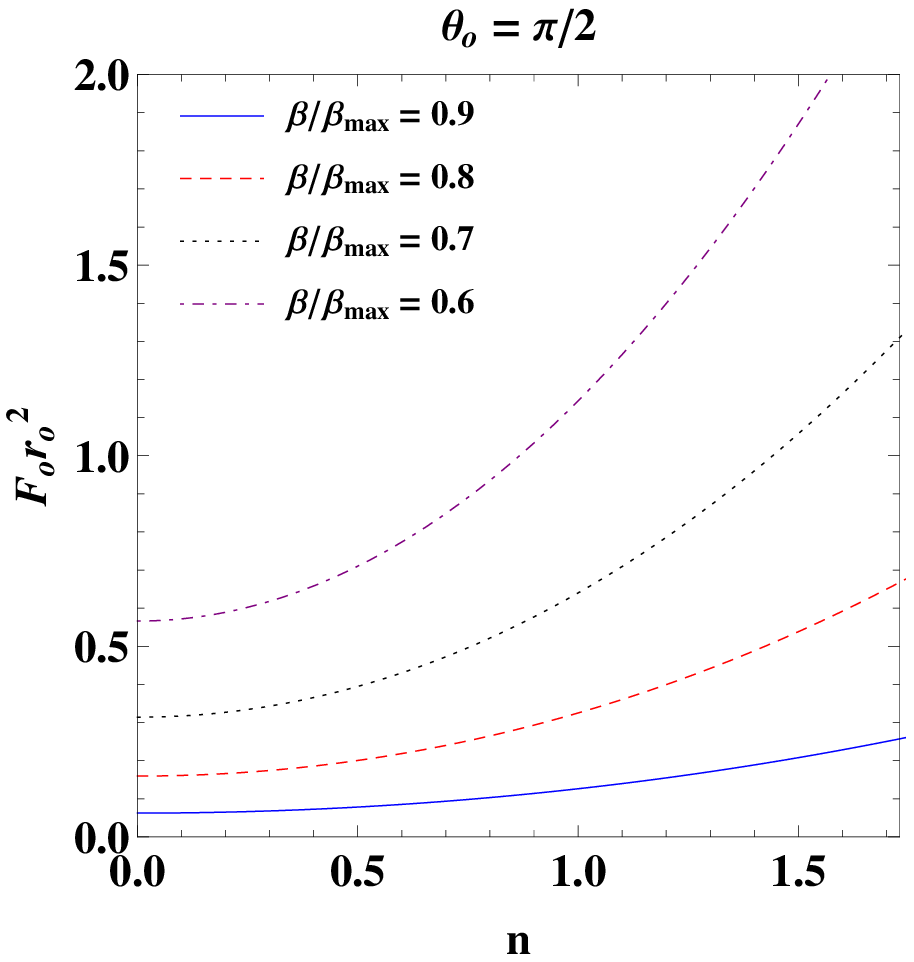}
\includegraphics[width=4cm]{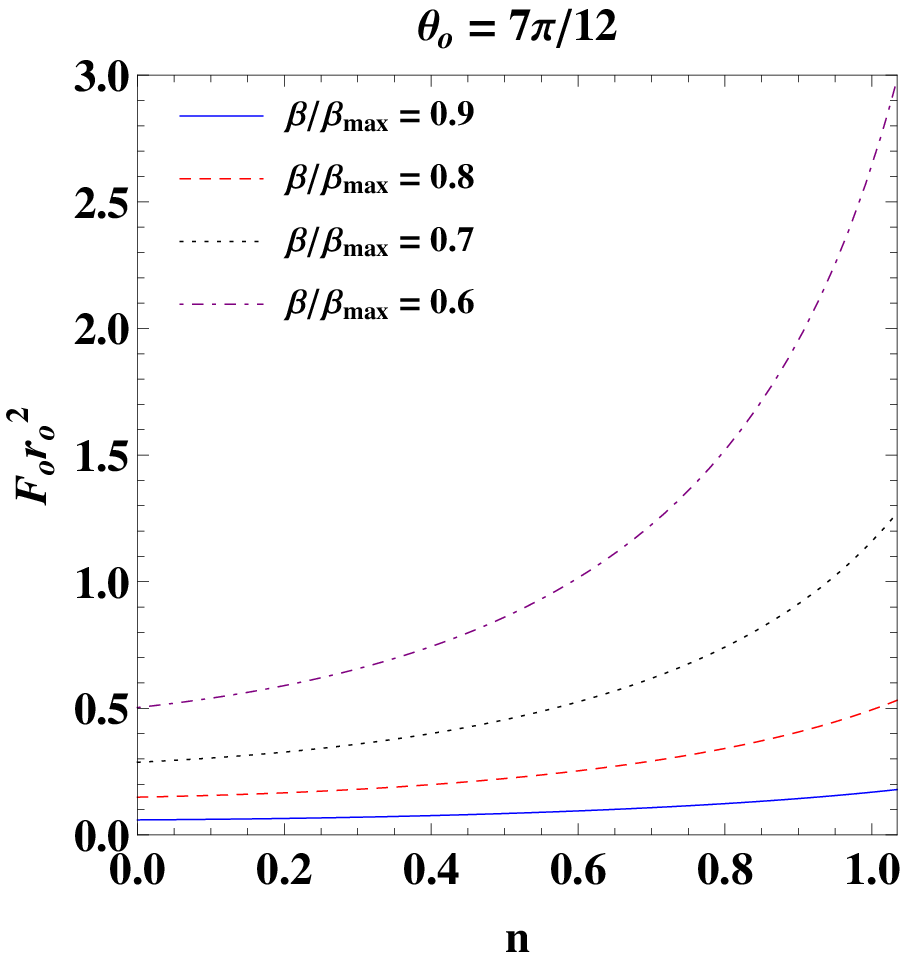}
\caption{The change of the flux $F_o$ at observer with the NUT charge $n$ for different $\beta/\beta_{max}$ and the observer's angular position $\theta_o$ as the source lies in the NHEKTN line. Here, we set $M=1$.}
\label{fig1s}
\end{figure}
\begin{figure}[ht]
\center
\includegraphics[width=5.2cm]{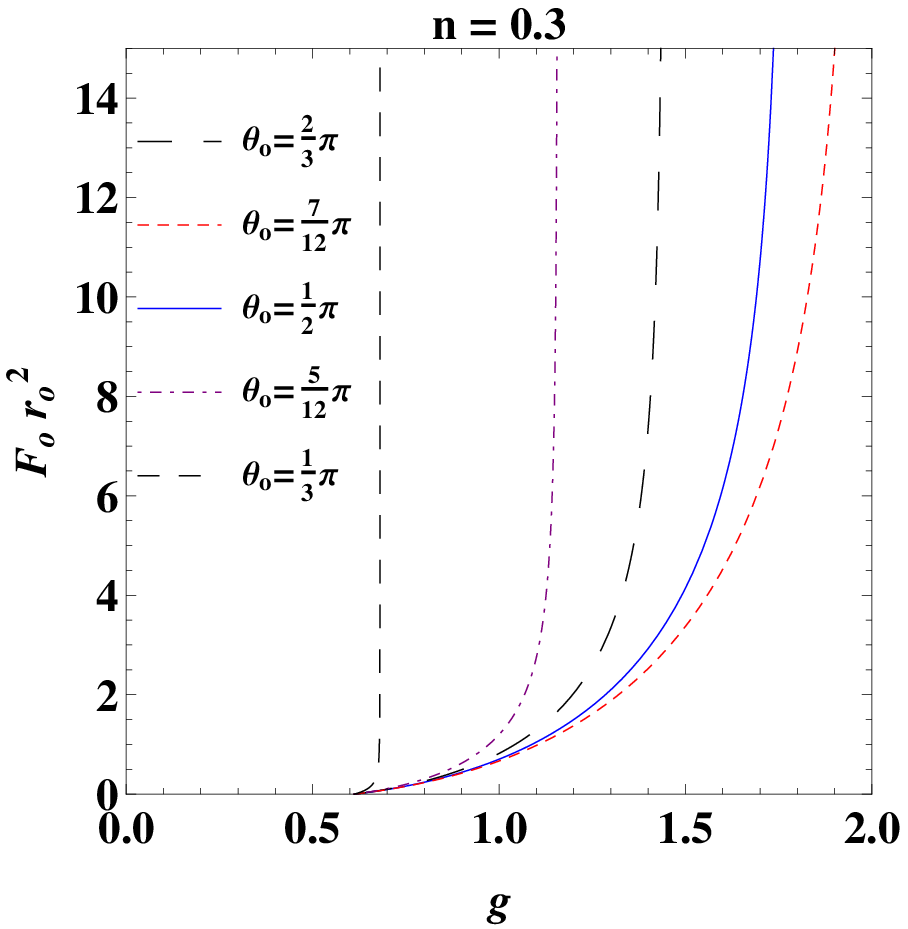}
\includegraphics[width=5cm]{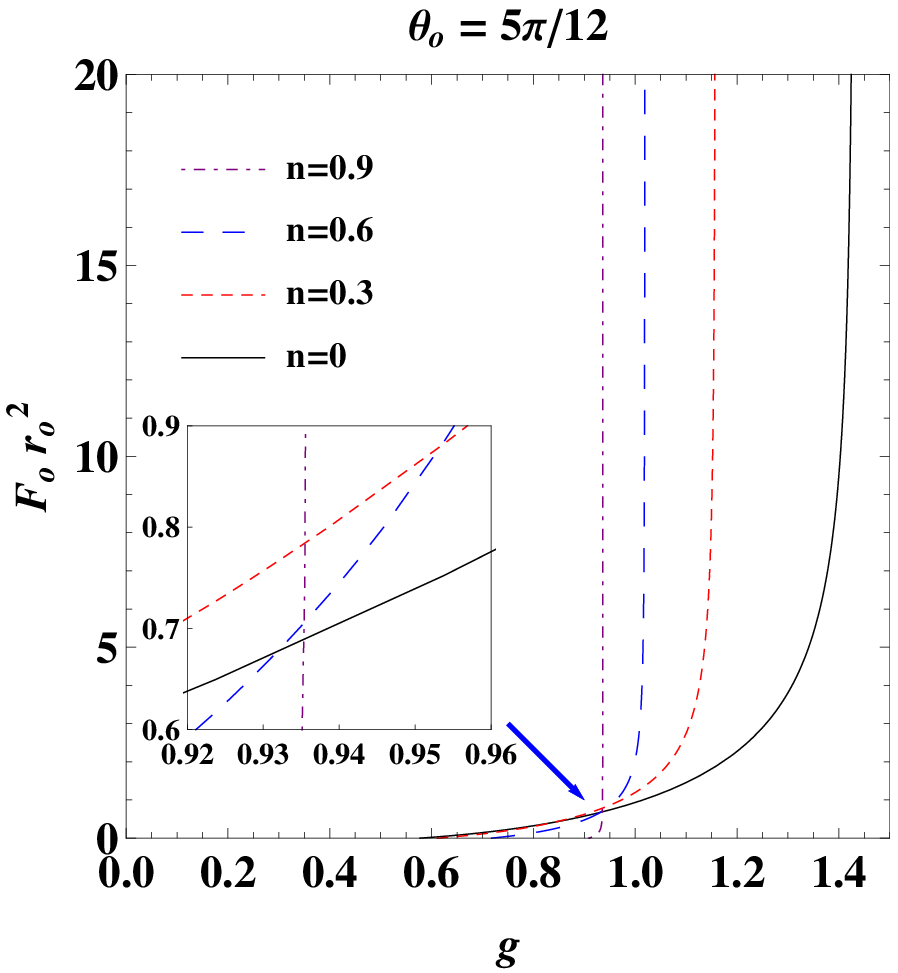}
\includegraphics[width=5cm]{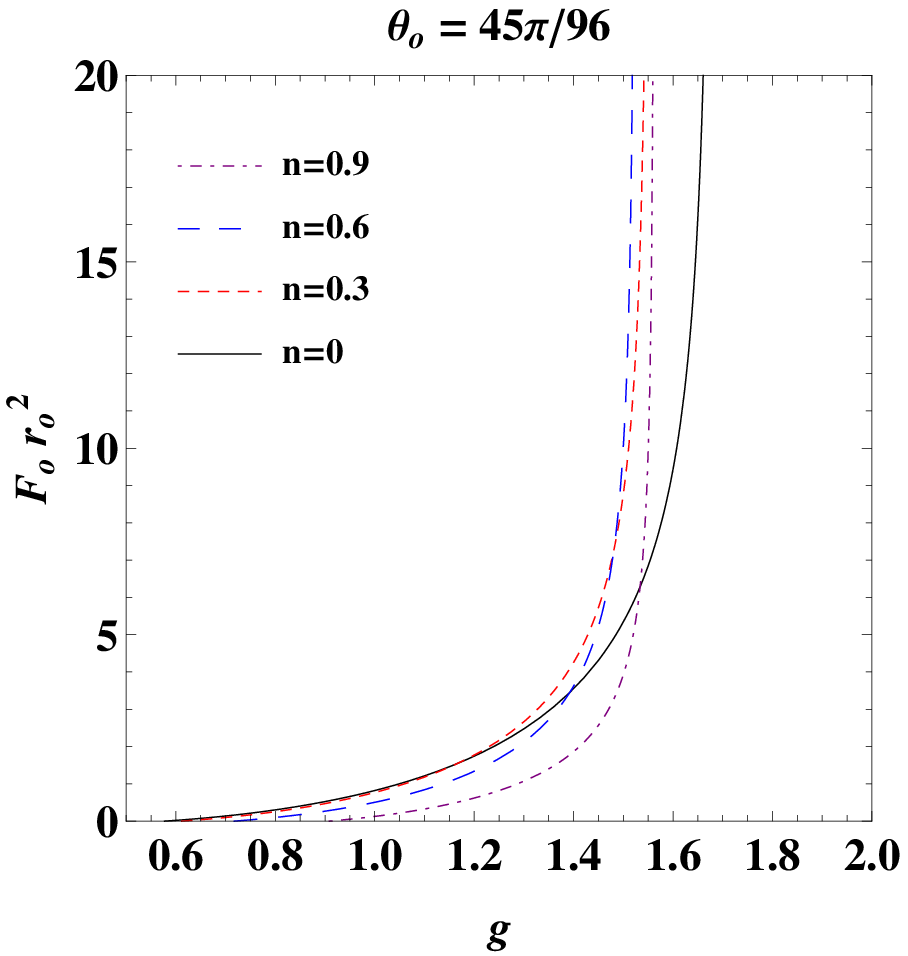}\\
\includegraphics[width=5cm]{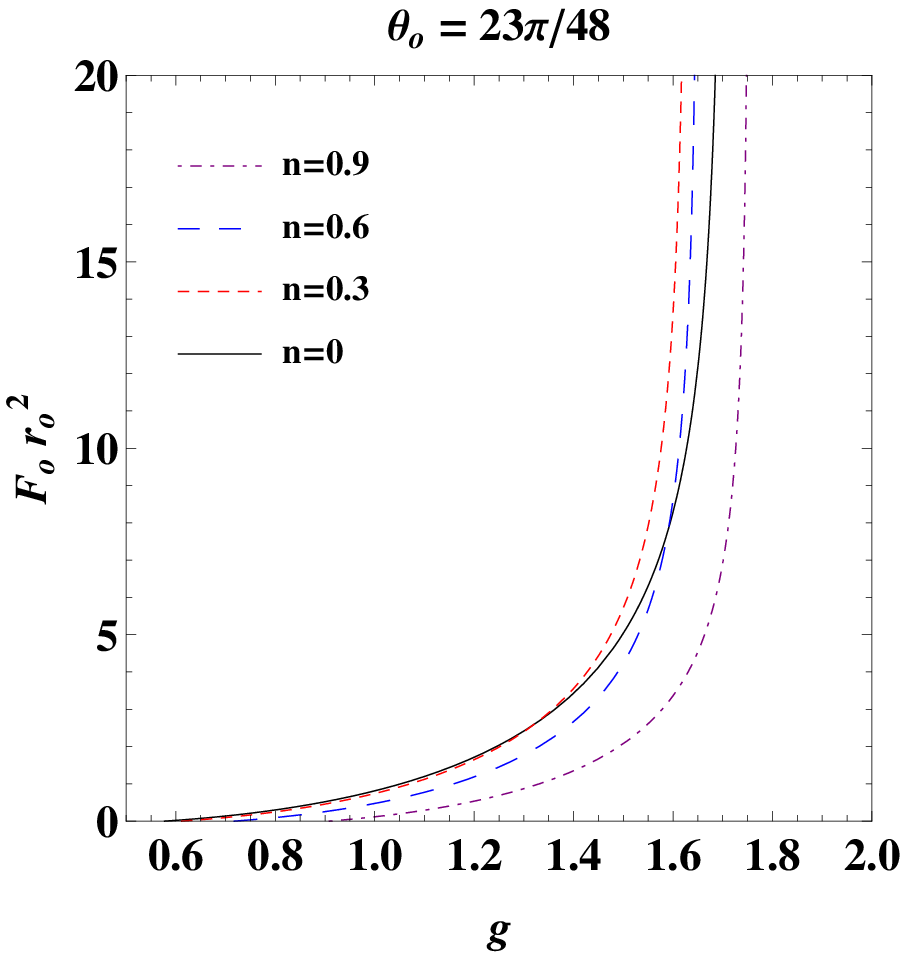}
\includegraphics[width=5cm]{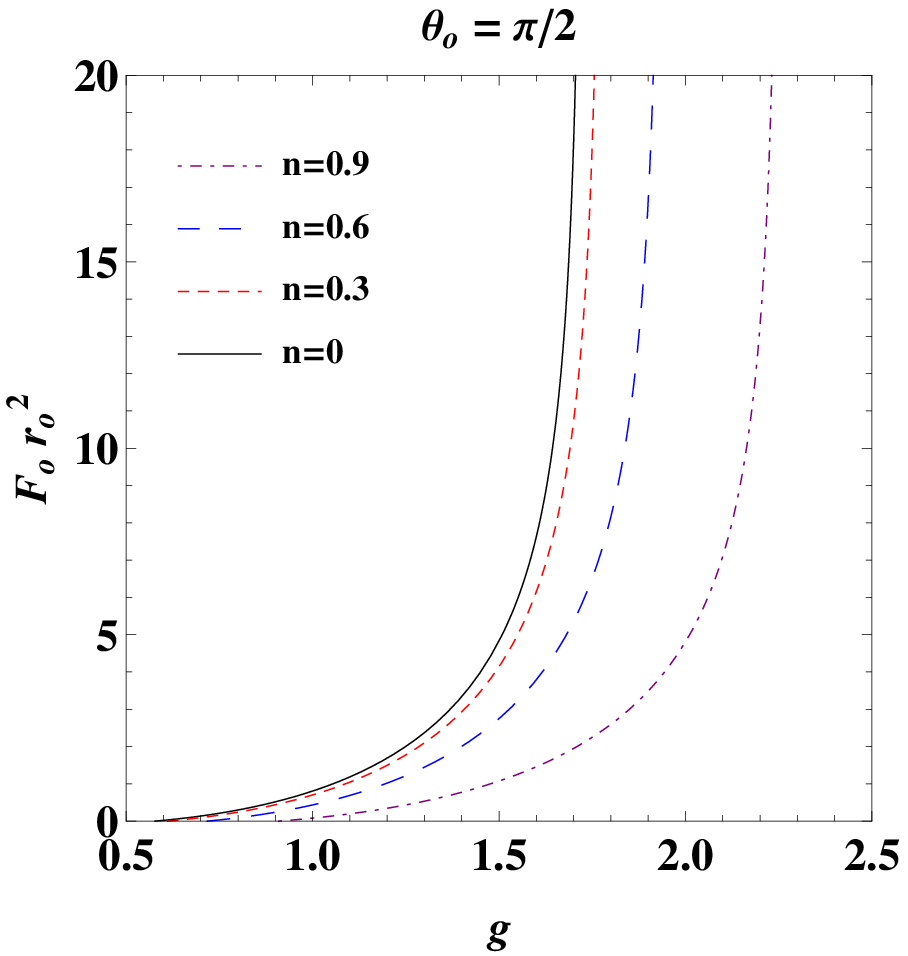}
\includegraphics[width=5cm]{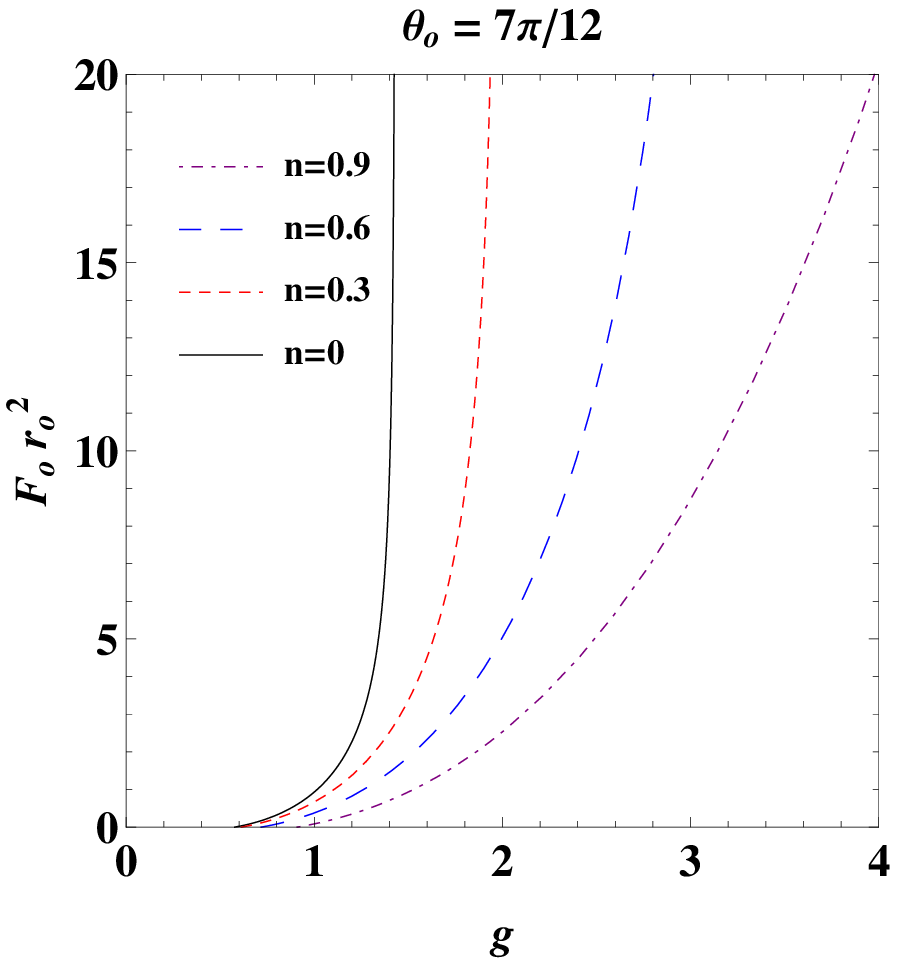}
\caption{The change of the flux $F_o$ at observer with the redshift factor $g$ for different NUT charge $n$ and the observer's angular position $\theta_o$ as the source lies in the NHEKTN line. Here, we set $M=1$.}
\label{fig2}
\end{figure}

For the observer with the fixed angular coordinate $\theta_o$, we find from Fig.(\ref{fig1s}) that the change of the flux $F_o$ with the NUT charge $n$ also depends on the coordinate $\theta_o$ and  the ratio $\beta/\beta_{max}$. For the case with $\theta_o\geq\pi/2$, the flux $F_o$ increases monotonously with $n$ for arbitrary fixed ratio $\beta/\beta_{max}$.
As $\theta_o=45\pi/96$, the flux $F_o$ decreases monotonously with $n$ for the smaller ratio $\beta/\beta_{max}$, but first increases and then decreases for the lager ratio $\beta/\beta_{max}$. With the further increase of the deviation quantity $\Delta\theta_o=\pi/2-\theta_o$ in the Northern Hemisphere
(for example, $\theta_o=5\pi/12$), the flux $F_o$ becomes a decreasing monotonously function of $n$ for arbitrary ratio $\beta/\beta_{max}$.  This means that the electromagnetic line emission from the near-horizon of extremal KTN black hole is brighter than that in the case of Kerr black hole for the observer in the Southern Hemisphere, but it becomes more faint as the observer's position deviates far from the equatorial plane in the Northern Hemisphere.  Moreover, in Fig. (\ref{fig1s}) we also note that in the cases with $\theta_o$ less than $\pi/2$ the flux vanishes for large enough NUT charge $n$.
\begin{figure}
\center
\includegraphics[width=4cm]{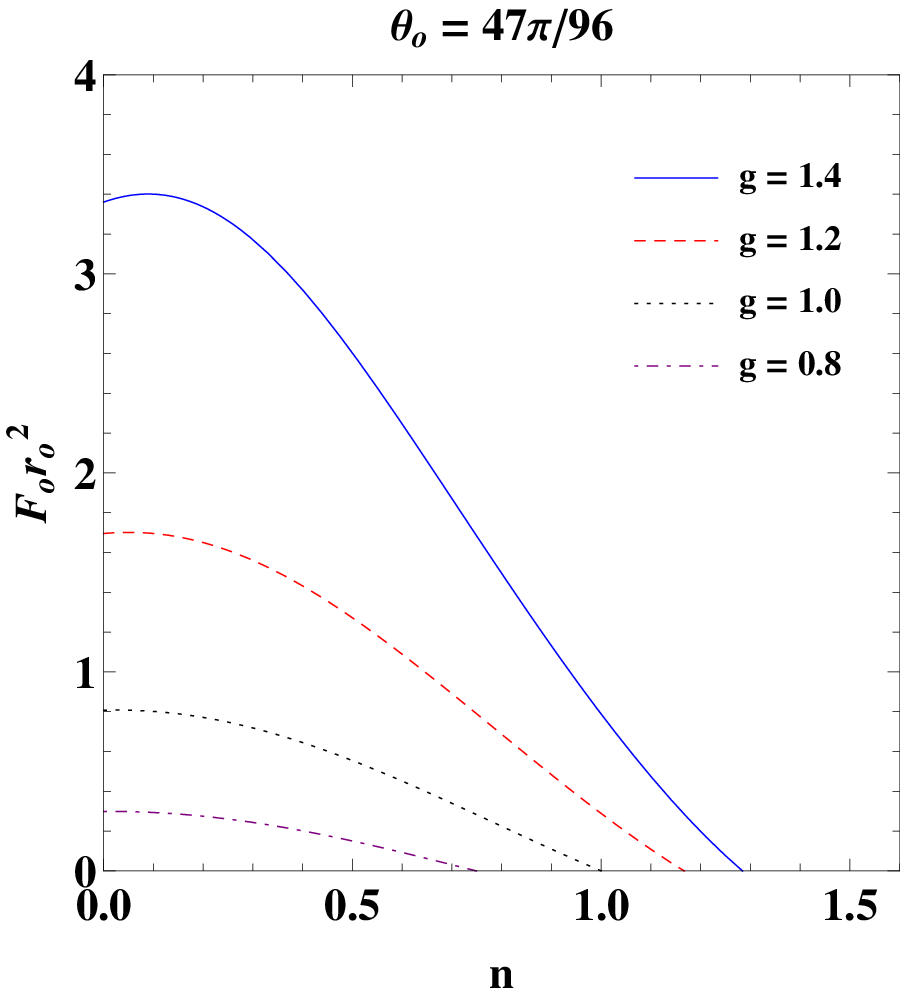}
\includegraphics[width=4cm]{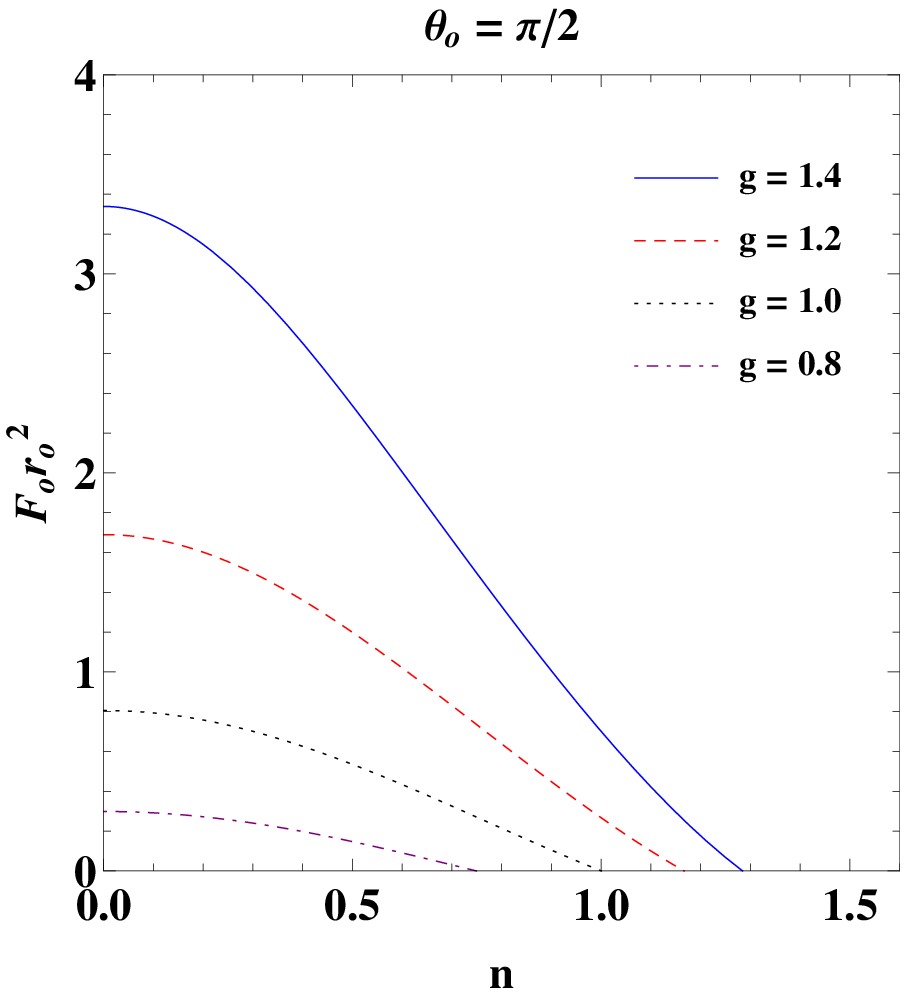}
\includegraphics[width=4cm]{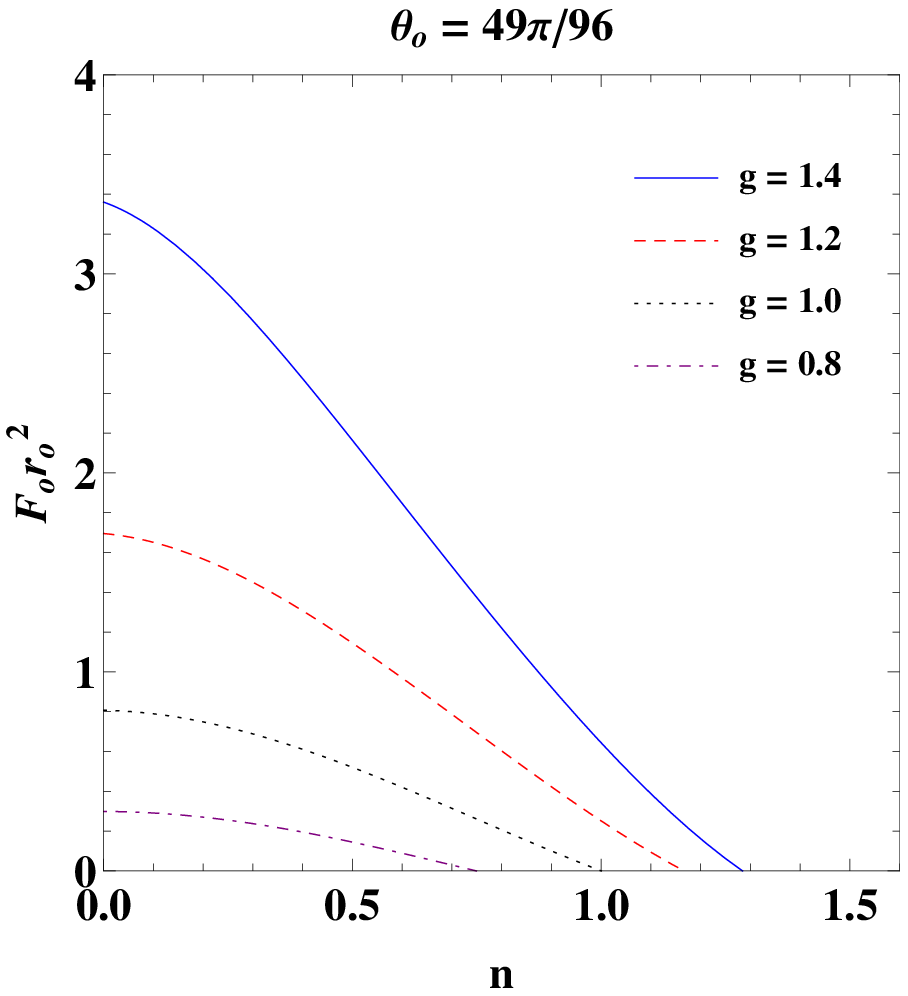}
\includegraphics[width=4cm]{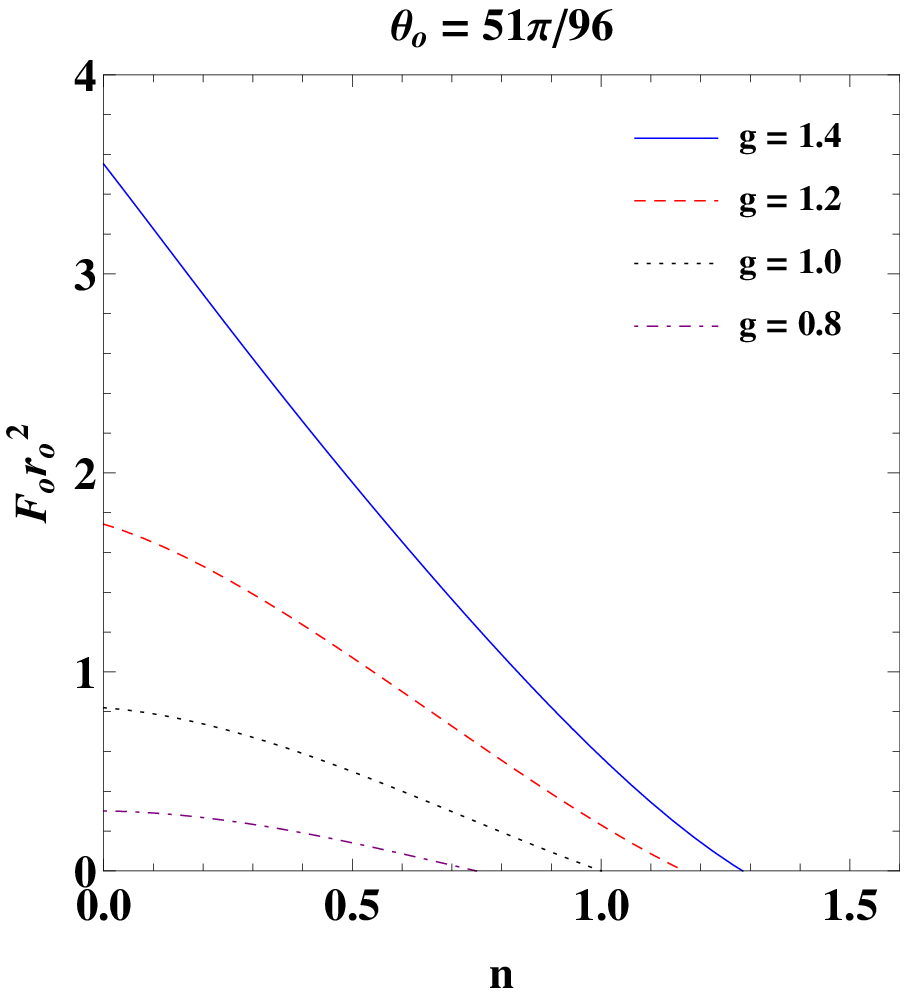}\\
\includegraphics[width=4.1cm]{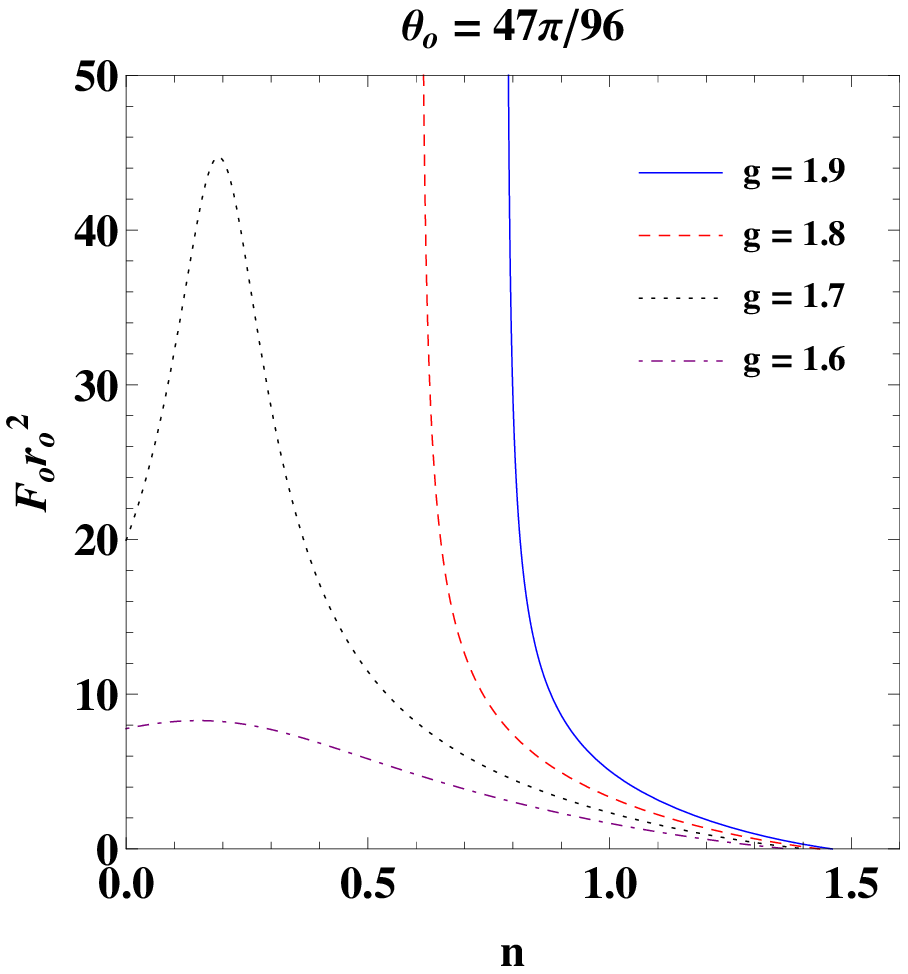}
\includegraphics[width=4.1cm]{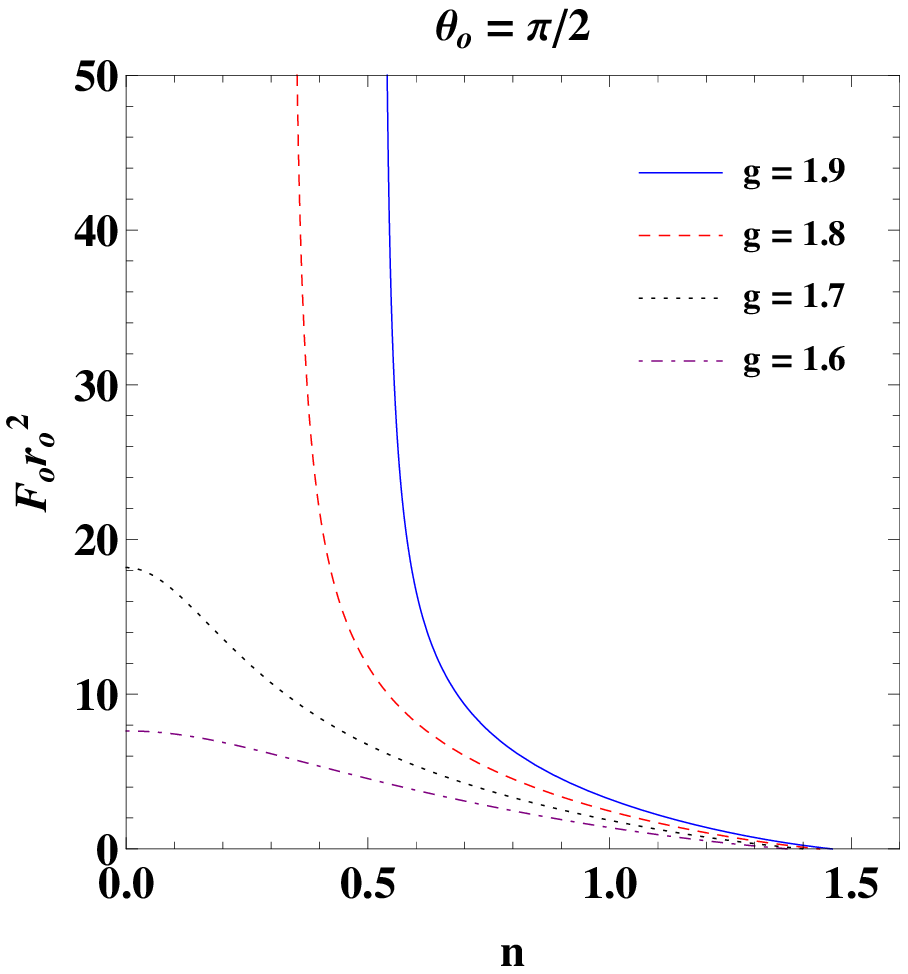}
\includegraphics[width=4.1cm]{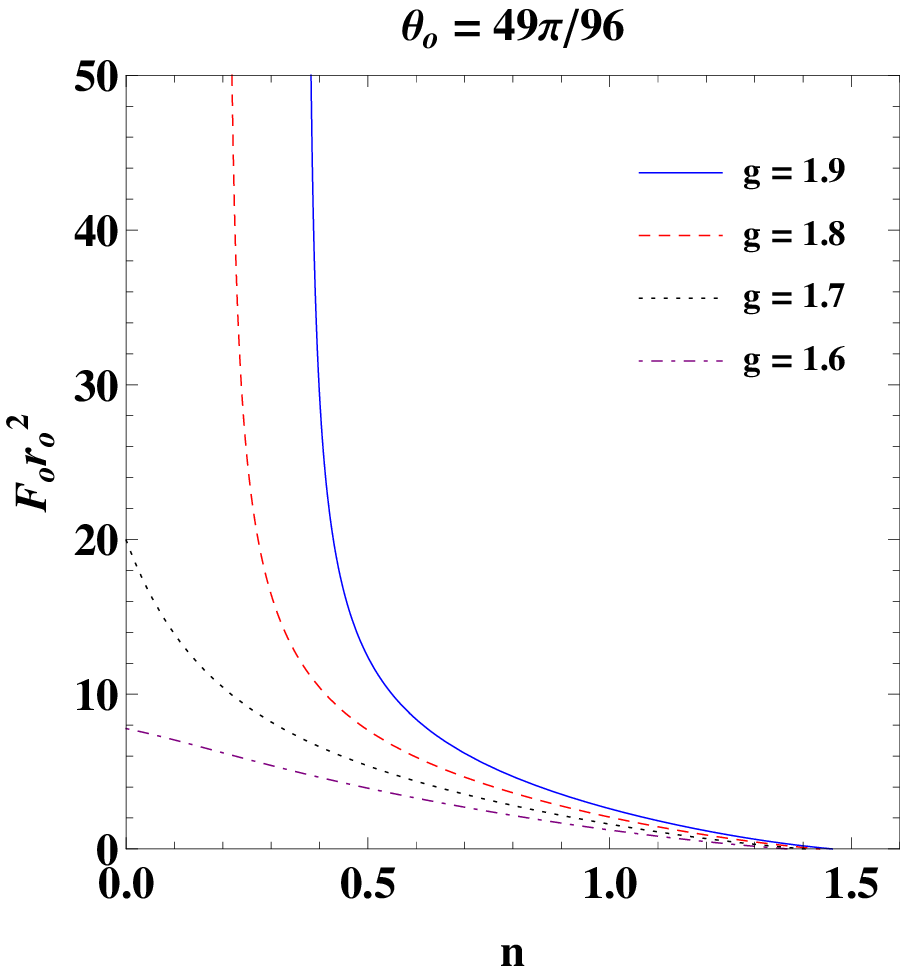}
\includegraphics[width=4.1cm]{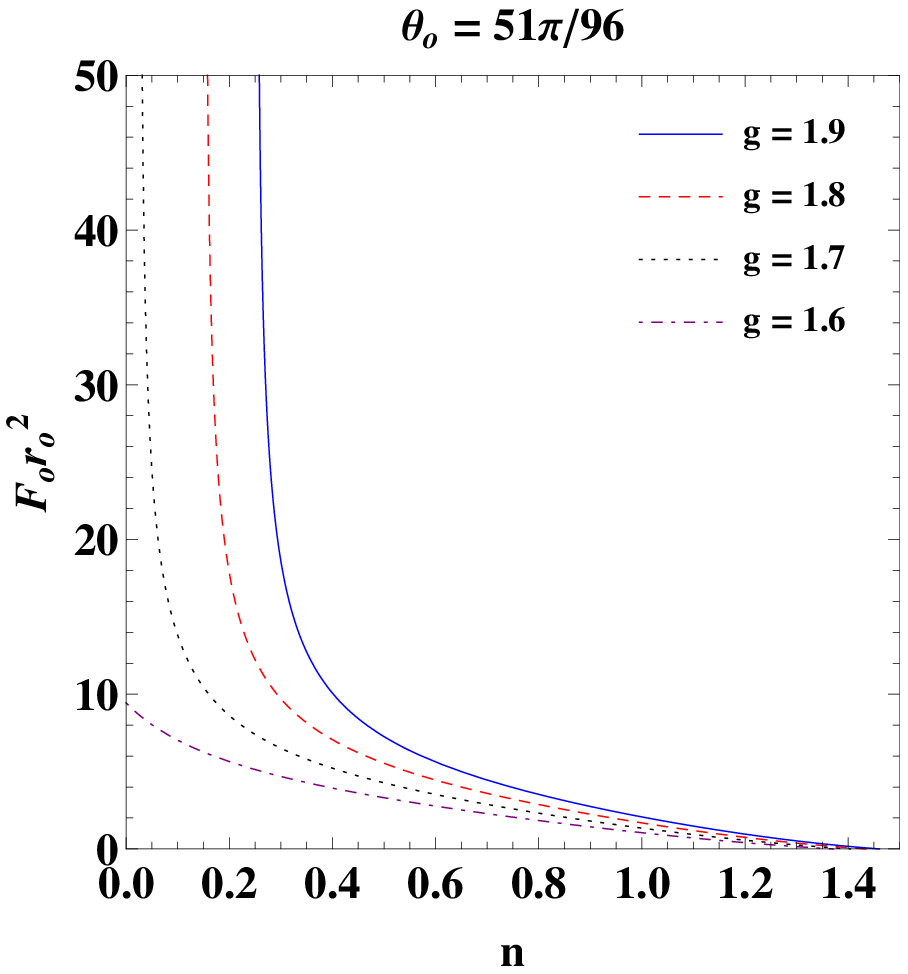}
\caption{The change of the flux $F_o$ at observer with the NUT charge $n$ for different redshift factor $g$ and the observer's angular position $\theta_o$ as the source lies in the NHEKTN line. Here, we set $M=1$.}
\label{fig21}
\end{figure}
\begin{figure}
\center
\includegraphics[width=4cm]{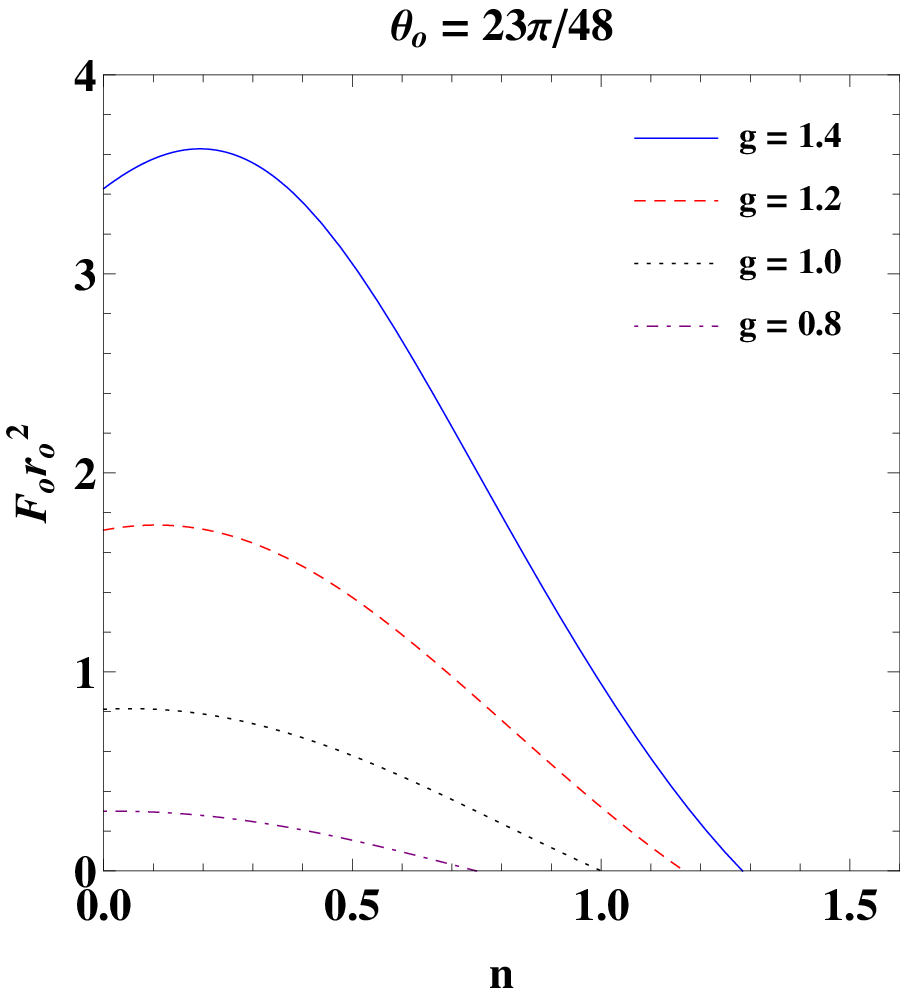}
\includegraphics[width=4.05cm]{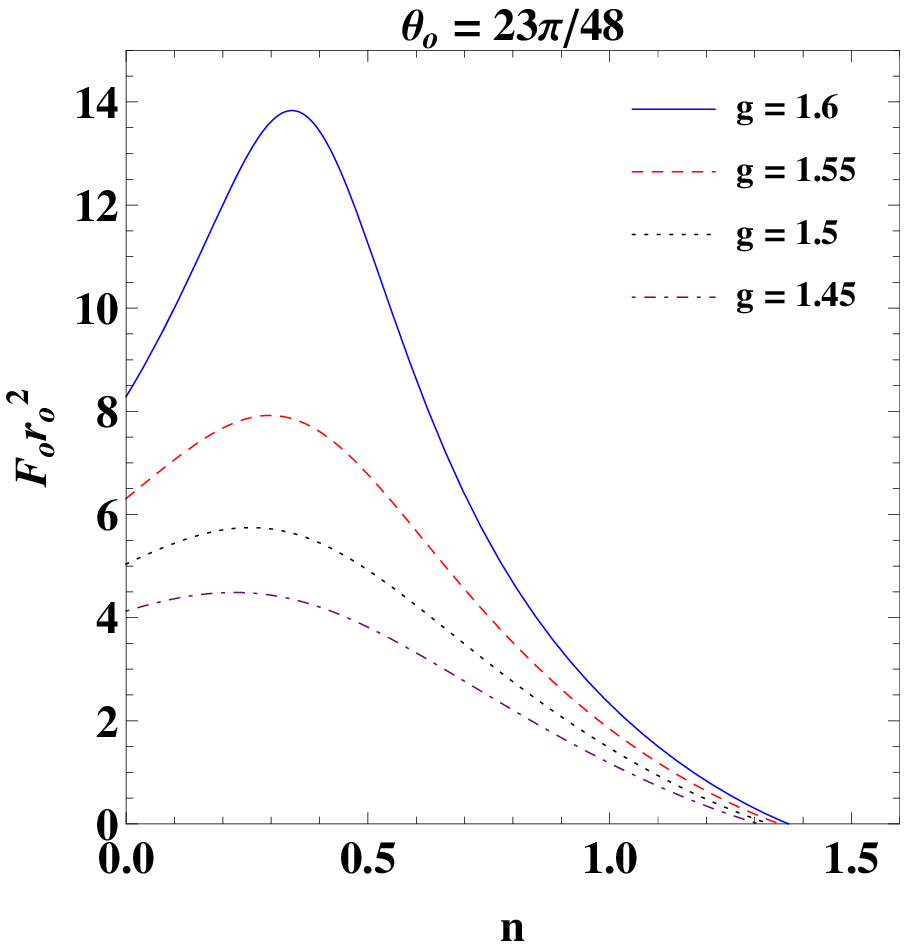}
\includegraphics[width=4.2cm]{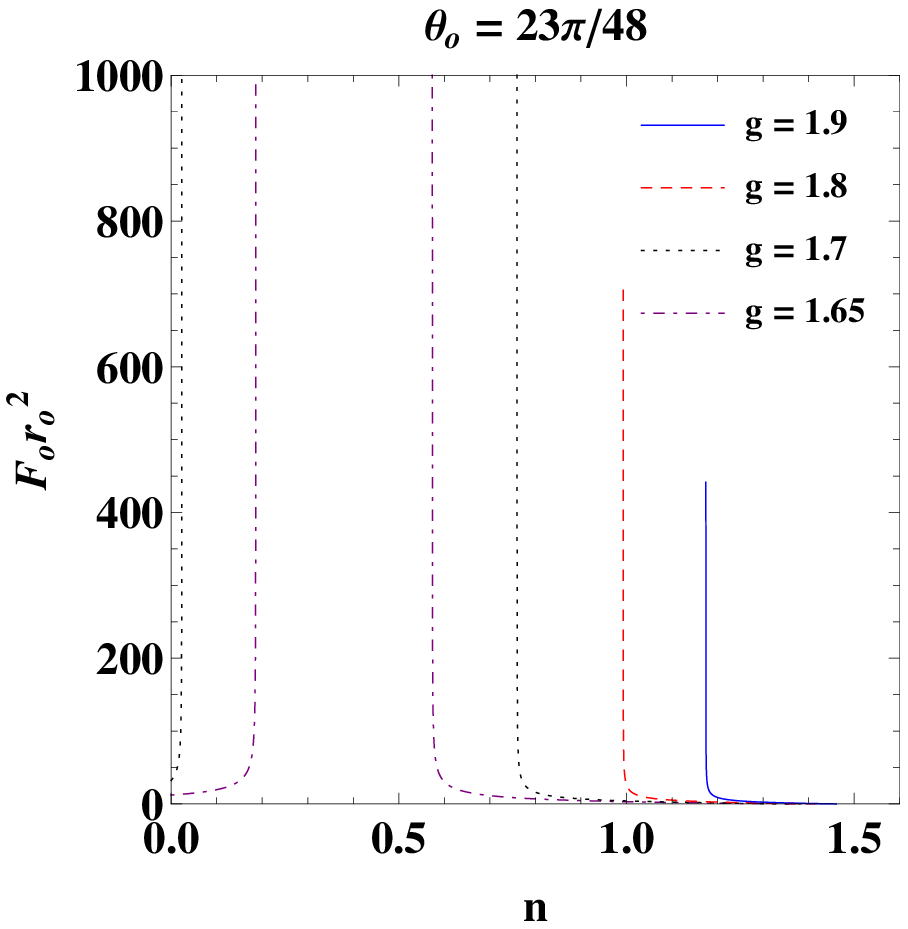}\\
\includegraphics[width=4.3cm]{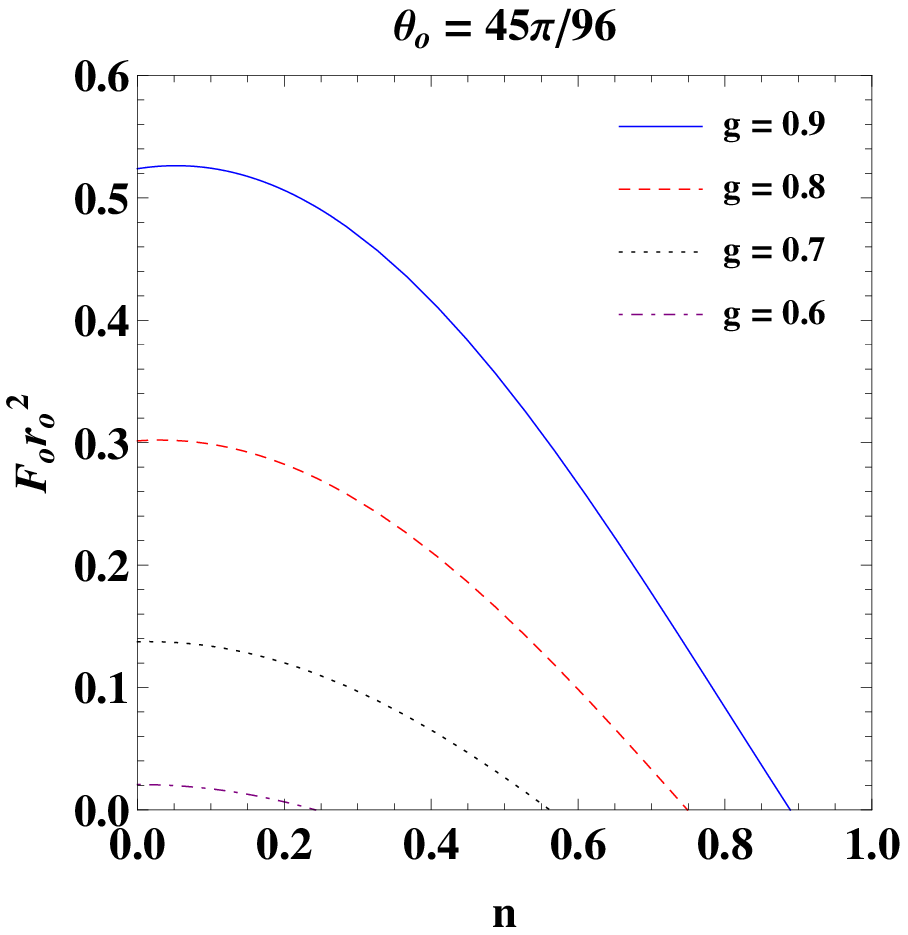}
\includegraphics[width=4cm]{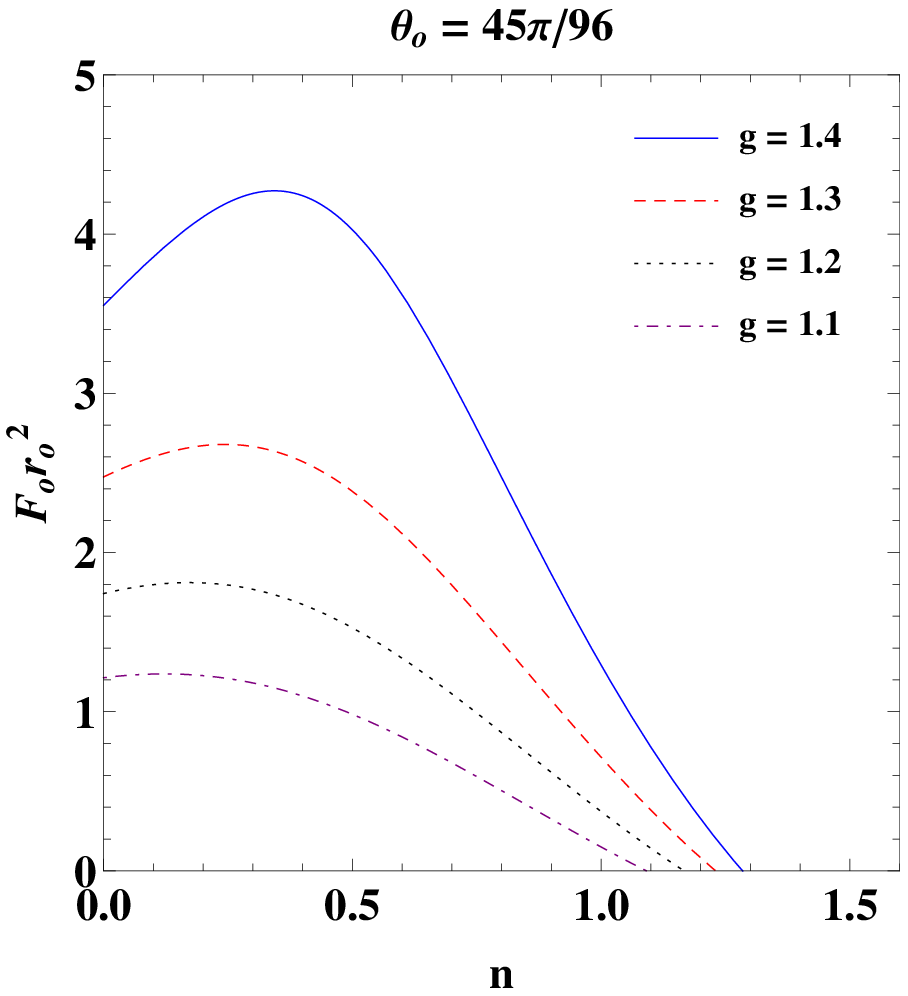}
\includegraphics[width=4.2cm]{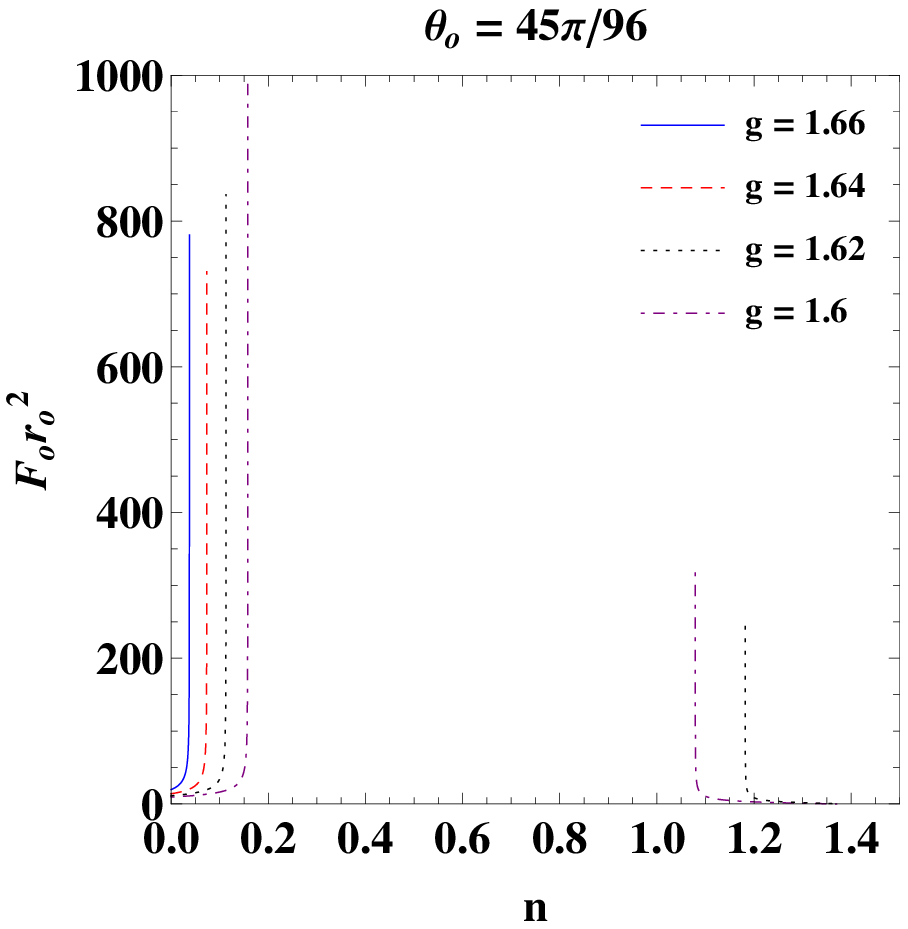}\\
\includegraphics[width=4.3cm]{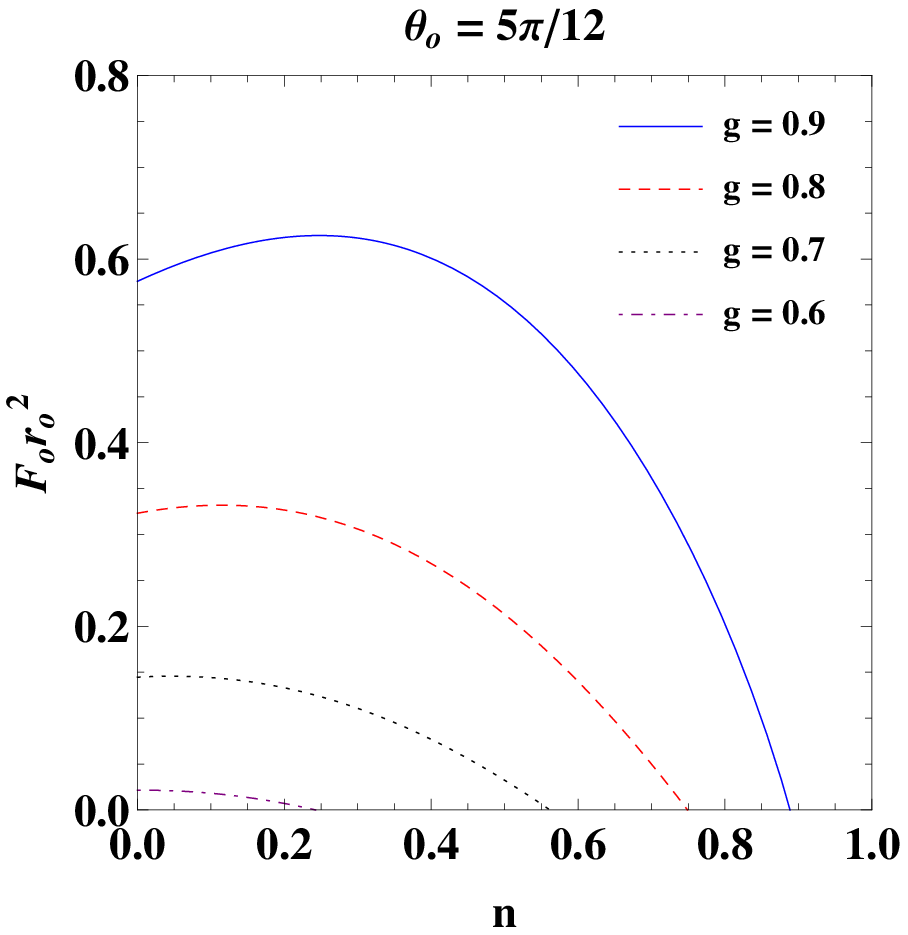}
\includegraphics[width=4.1cm]{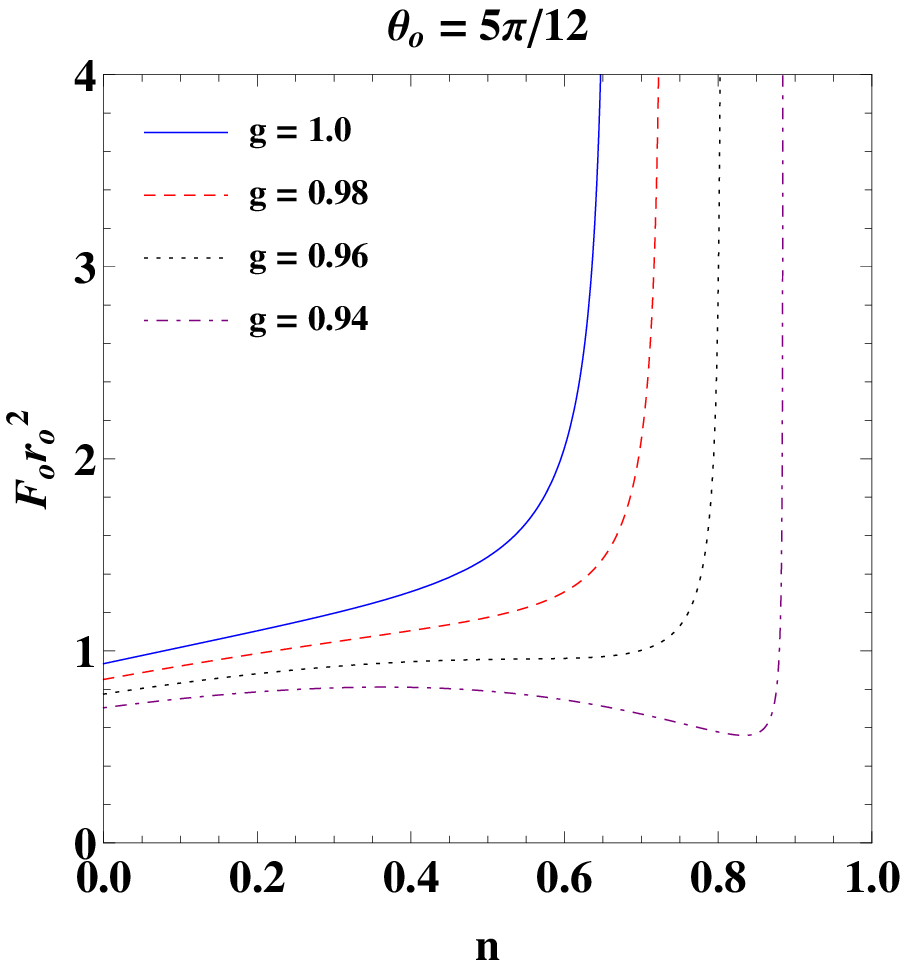}
\includegraphics[width=4.2cm]{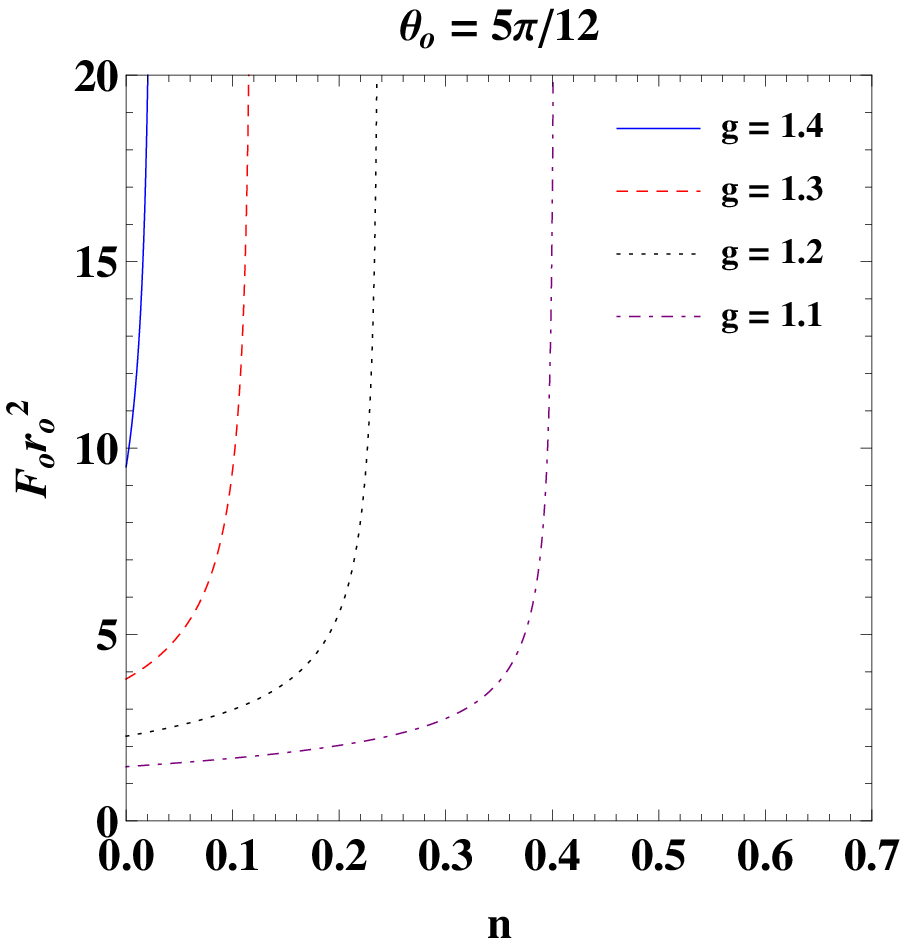}\\
\caption{The change of the flux $F_o$ at observer with the NUT charge $n$ for different redshift factor $g$ and the observer's angular position $\theta_o$ as the source lies in the NHEKTN line. Here, we set $M=1$.}
\label{fig22}
\end{figure}
In Fig.(\ref{fig2}), we plot the dependence of the flux $F_o$ of electromagnetic emission from  NHEKTN line on the redshift factor $g$ for different NUT charge $n$ and the observer's angular position $\theta_o$.
Fig.(\ref{fig2}) indicates that the flux $F_o$ increases with the redshift factor $g$ for fixed $n$ and $\theta_o$, which could be explained by a fact from Eq. (\ref{gred}) that the larger value of $g$ means that the higher energy at observer $E_o$ can be obtained for the fixed  energy $E_s$ of  photons at the source. This behavior of $F_o$ is the same as that in the Kerr case \cite{P19}. Figs.(\ref{fig21}) and (\ref{fig22}) tell us that the change of the flux $F_o$ with the NUT charge $n$ for the fixed redshift factor $g$ is more complicated, which depends also on the observer's angular position $\theta_o$. We find that the flux $F_o$ exists only
in the case where $g$ is in the certain range of $(g_{c1}, g_{c2})$, and the values of $g_{c1}$ and $g_{c2}$ depend on the NUT charge $n$ and the observer's position $\theta_o$. This is consistent with the previous discussion from Eq. (\ref{gcc}). Moreover,  the range of $g$ with the non-zero flux $F_o$ becomes narrow with increase of the deviation of observer from the equatorial plane in the Northern Hemisphere.
As the observer lies in the equatorial plane or in the Southern Hemisphere, one can find that the flux $F_o$ decreases monotonously with $n$ for fixed redshit $g$.
As the observer deviates from the equatorial plane in the Northern Hemisphere, we find that the flux $F_o$ still decreases with $n$ in the cases with the lower redshift factor $g$ or the larger one, but it first increases and then decreases with $n$ for the case with the intermediate redshift factor $g$. With the further deviation for observer, the region of redshift factor $g$ for the flux decreasing with NUT charge becomes gradually narrow. Finally, we find that the region of the flux decreasing with $n$ as $\theta_o=5\pi/12$ vanishes in the case with the larger redshift factor $g$. Thus, the NUT charge modifies the electromagnetic line emissions from near-horizon region of an extremal KTN black hole.

\section{Summary}

We have studied electromagnetic line emissions from near-horizon region in the extremal KTN black hole spacetime. Our result show that the fluxes $F_o$ of electromagnetic line emissions from near-horizon region depend on the NUT charge $n$, the positions of source and observer, and the redshift factor $g$. For the fixed NUT charge $n$, the flux $F_o$ decreases monotonously with the ratio $\beta/\beta_{max}$.  This means that the brightest electromagnetic line at observer is emitted from the sources located in the equatorial plane. This property of electromagnetic emissions in NHEKTN line is similar to those in the Kerr case. However, for the fixed NUT charge $n$, one can find  that the flux $F_o$ increases with the angular coordinate $\theta_o$ of the observer's position, which implies that electromagnetic line emission for the observer in the Southern Hemisphere is brighter than that in the Northern one. This is different from that in the Kerr case where the electromagnetic line emission is the brightest for the observer in the equatorial plane. For the observer with the fixed angular coordinate $\theta_o$, we find that the electromagnetic line emission from the near-horizon of an extremal KTN black hole is brighter than that in the case of Kerr black hole for the observer in the equatorial plane or in the Southern Hemisphere, but it becomes more faint as the observer's position deviates far from the equatorial plane in the Northern Hemisphere.

We also study the dependence of the flux $F_o$ of electromagnetic emission from  NHEKTN line on the redshift factor $g$ for different NUT charge $n$ and the observer's angular position $\theta_o$.
With the increase of redshift factor $g$,  the flux $F_o$ increases for fixed $n$ and $\theta_o$, which is the similar as that in the Kerr case \cite{P19}. However, for the fixed redshift factor $g$, the change of the flux $F_o$ with the NUT charge $n$  is more complicated. Firstly, the flux $F_o$ exists only
in the case where $g$ is in the certain range of $(g_{c1}, g_{c2})$, and the values of $g_{c1}$ and $g_{c2}$ depend on the NUT charge $n$ and the observer's position $\theta_o$. The range of $g$ with the non-zero flux $F_o$ becomes narrow with the deviation of observer from the equatorial plane in the Northern Hemisphere.
As the observer lies in the equatorial plane or in the Southern Hemisphere, the flux $F_o$ decreases monotonously with $n$ for fixed redshit $g$. As the observer deviates from the equatorial plane in the Northern Hemisphere, the flux $F_o$ still decreases with $n$ in the cases with the lower redshift factor $g$ or the larger one, but it first increases and then decreases with $n$ for the case with the intermediate redshift factor $g$. With the further deviation of observer in the Northern Hemisphere, the region of redshift factor $g$ for the flux decreasing with NUT charge becomes gradually narrow. Thus, the NUT charge modifies the electromagnetic line emissions from near-horizon region of an extremal KTN black hole.

\section{\bf Acknowledgments}

This work was partially supported by the National Natural Science Foundation of China under
Grant No. 11875026, the Scientific Research
Fund of Hunan Provincial Education Department Grant
No. 17A124. J. Jing's work was partially supported by
the National Natural Science Foundation of China under
Grant No. 11475061, 11875025.

\begin{appendices}
\section{Appendix}
We now derivate the celestial coordinate $(\alpha,\beta)$ (\ref{tianq}) of the image pixel on the observer's local sky in the KTN spacetime. We assume that the observer is located in the spatial infinity in a KTN black hole spacetime.
As $\hat{r} \rightarrow \infty$, the metric (\ref{metric1}) becomes
\begin{eqnarray}\label{apm1}
ds^2&=&-(dt+2n\cos{\theta}d\phi)^2+d\hat{r}^2+\hat{r}^2d\theta^2+\hat{r}^2\sin^2{\theta}d\phi^2,
\end{eqnarray}
which is not asymptotically flat due to the existence of the cross term $dtd\phi$. However,
setteing $dt'=dt+2n\cos{\theta}d\phi$ and making the coordinate transformation
\begin{eqnarray}
x=\rho\sin\theta\cos\phi,\quad\quad\quad\quad y=\rho\cos\theta, \quad\quad\quad\quad z=\rho\sin\theta\sin\phi,
\end{eqnarray}
we find that the above metric (\ref{apm1}) can be rewritten as
\begin{eqnarray}
ds^2&=&-dt'^2+dx^2+dy^2+dz^2.
\end{eqnarray}
Thus, in the KTN black hole spacetime, the observer basis at the spatial infinity $\left\{e_{\tilde{t'}}, e_{\tilde{x}}, e_{\tilde{y}}, e_{\tilde{z}}\right\}$  can be expanded as a form in the coordinate basis $\left\{\partial_{t}, \partial_{\hat{r}}, \partial_{\theta}, \partial_{\phi}\right\}$  \cite{P28}
\begin{eqnarray}
e_{\tilde{\mu}}=e_{\tilde{\mu}}^{\nu} \partial_{\nu}.\label{coordinate1}
\end{eqnarray}
Here  the transform matrix $e^{\nu}_{\tilde{\mu}}$ satisfies $g_{\mu\nu}e^{\mu}_{\tilde{\alpha}}e^{\nu}_{\tilde{\beta}}=\eta_{\tilde{\alpha}\tilde{\beta}}$ and $\eta_{\tilde{\alpha}\tilde{\beta}}$ is the  Minkowski metric. As in Ref. \cite{P28}, we can chose that the transform matrix $e^{\nu}_{\tilde{\mu}}$  has a form
\begin{eqnarray}
e_{\tilde{\mu}}^{\nu}=\left( \begin{array}{ccccc}
{\zeta} & {0}           & {0}          & {\gamma}  \\
{0}     & {A^{\hat{r}}} & {0}          & {0}    \\
{0}     & {0}           & {A^{\theta}} & {0}  \\
{0}     & {0}           & {0}          & {A^{\phi}}  \end{array}\right),\label{coordinate3}
\end{eqnarray}
where $\zeta$,  $\gamma$, $A^{\hat{r}}$, $A^{\theta}$,  and   $A^{\phi}$ are real coefficients. From the Minkowski normalization
\begin{eqnarray}
e_{\tilde{\mu}}e^{\tilde{\nu}}=\delta_{\tilde{\mu}}^{\;\tilde{\nu}},
\end{eqnarray}
one can obtain
\begin{eqnarray}
\begin{aligned} A^{\hat{r}} &=\frac{1}{\sqrt{g_{\hat{r}\hat{r}}}}, \qquad A^{\theta}=\frac{1}{\sqrt{g_{\theta \theta} }}, \qquad A^{\phi}=\frac{1}{\sqrt{g_{\phi \phi}}}, \\ \zeta &=\sqrt{\frac{g_{\phi \phi}}{g_{t \phi}^{2}-g_{t t} g_{\phi \phi}}}, \qquad \gamma=-\frac{g_{t \phi}}{g_{\phi \phi}} \sqrt{\frac{g_{\phi \phi}}{g_{t \phi}^{2}-g_{t t} g_{\phi \phi}}}. \end{aligned}
\end{eqnarray}
The locally measured four-momentum $p^{\tilde{\mu}}$ of a photon can be obtained by the projection of its four-momentum $p^{\mu}$  onto $e_{\tilde{\mu}}$, i.e., $p^{\tilde{\mu}}=\eta^{\tilde{\mu}\tilde{\nu}}p_{\tilde{\nu}}=\eta^{\tilde{\mu}\tilde{\nu}}e_{\tilde{\nu}}^{\nu} p_{\nu}$. This means that  $p^{\tilde{\mu}}$ has the form
\begin{eqnarray}
\begin{aligned}p^{\tilde{t'}}=\zeta E-\gamma p_{\phi},\qquad \qquad  p^{\tilde{x}}=\frac{1}{\sqrt{g_{\hat{r} \hat{r}}}}p_{\hat{r}},\\ p^{\tilde{y}}=\frac{1}{\sqrt{g_{\theta \theta}}} p_{\theta}, \qquad \qquad  p^{\tilde{z}}=\frac{1}{\sqrt{g_{\phi \phi}}} p_{\phi}.
\end{aligned}
\end{eqnarray}
In the orthonormal basis $\left\{e_{\tilde{x}}, e_{\tilde{y}}, e_{\tilde{z}}\right\}$, the 3-vector linear momentum $\vec{p}$  of the photon can be expressed in terms of the components $p_{\tilde{x}},\ p_{\tilde{y}}$ and $p_{\tilde{z}}$, i.e.,
\begin{eqnarray}
\vec{p}&=p^{\tilde{x}} e_{\tilde{x}}+p^{\tilde{y}} e_{\tilde{y}}+p^{\tilde{z}}e_{\tilde{z}}.
\end{eqnarray}
From the geometry of the photon's detection, we have
\begin{eqnarray}
p^{\tilde{x}} =|\vec{p}| \cos \varphi \cos \psi, \quad\quad\quad
p^{\tilde{y}} =|\vec{p}| \sin \varphi,\quad\quad\quad
p^{\tilde{z}} =|\vec{p}| \cos \varphi \sin \psi,
\end{eqnarray}
and then the celestial coordinates $(\alpha,\beta)$ of the image pixel on observer's sky become
\begin{eqnarray}
\alpha &=&-\lim _{\hat{r}_{o} \rightarrow \infty} \hat{r}_{o}\tan \psi=-\lim _{\hat{r}_{o} \rightarrow \infty}\hat{r}_{o} \frac{p^{\tilde{z}}}{p^{\tilde{x}}}=-\frac{\hat{\lambda}}{\sin{\theta_o}},\nonumber\\
\beta &=&\lim _{\hat{r}_{o} \rightarrow \infty} \hat{r}_{o}\frac{\tan \varphi}{\cos{ \psi}}=\lim _{\hat{r}_{o} \rightarrow \infty} \hat{r}_{o} \frac{p^{\tilde{y}}}{p^{\tilde{x}}}=\pm\sqrt{\hat{\Theta}(\theta_o)},
\label{coordinate7}
\end{eqnarray}
for the light ray passing through a KTN black hole (\ref{metric1}), which is consistent with that in Refs.\cite{P28,KTNapp8}.
\end{appendices}

\end{document}